\documentclass[onecolumn, twocolumn, mathlines]{aastex631}
\usepackage{mathtools}
\usepackage{amsmath}
\usepackage{soul}
\usepackage{lipsum}
\usepackage{float}
\usepackage{booktabs}
\usepackage{rotating}
\usepackage{longtable}

\usepackage[T1]{fontenc} 
\usepackage[utf8]{inputenc} 

\newcommand\cms{$\mathrm{cm\ s^{-1}}$}
\newcommand\ms{$\mathrm{m\ s^{-1}}$}

\newcommand{\target}{HD 26965}

\newcommand{\PSUAA}{Department of Astronomy \& Astrophysics, 525 Davey Laboratory, The Pennsylvania State University, University Park, PA, 16802, USA}
\newcommand{\PSUCEHW}{Center for Exoplanets and Habitable Worlds, 525 Davey Laboratory, The Pennsylvania State University, University Park, PA, 16802, USA}
\newcommand{\PSETI}{Penn State Extraterrestrial Intelligence Center, 525 Davey Laboratory, The Pennsylvania State University, University Park, PA, 16802, USA}
\newcommand{\UA}{Steward Observatory, University of Arizona, 933 N.\ Cherry Ave, Tucson, AZ 85721, USA}

\newcommand{\Penn}{Department of Physics and Astronomy, University of Pennsylvania, 209 S 33rd St, Philadelphia, PA 19104, USA}

\newcommand{\STScI}{Space Telescope Science Institute, 3700 San Martin Dr, Baltimore, MD 21218, USA}
\newcommand{\JHU}{Department of Physics and Astronomy, Johns Hopkins University, 3400 N Charles St, Baltimore, MD 21218, USA}
\newcommand{\GoddardESAL}{Exoplanets and Stellar Astrophysics Laboratory, NASA Goddard Space Flight Center, Greenbelt, MD 20771, USA}

\newcommand{\Macquarie}{School of Mathematical and Physical Sciences, Macquarie University, Balaclava Road, North Ryde, NSW 2109, Australia
}

\newcommand{\JPL}{Jet Propulsion Laboratory, California Institute of Technology, 4800 Oak Grove Drive, Pasadena, California 91109}

\newcommand{\UCI}{Department of Physics \& Astronomy, The University of California, Irvine, Irvine, CA 92697, USA}
\newcommand{\Carleton}{Carleton College, One North College St., Northfield, MN 55057, USA}

\newcommand{\PSUICS}{Institute for Computational and Data Sciences, The Pennsylvania State University, University Park, PA, 16802, USA}
\newcommand{\PSUCASt}{Center for Astrostatistics, 525 Davey Laboratory, The Pennsylvania State University, University Park, PA, 16802, USA}

\shorttitle{Stellar activity in HD 26965}
\shortauthors{Burrows et al.}

\begin{document}

\title{The death of Vulcan: NEID reveals the planet candidate orbiting HD 26965\footnote{{\target} is also known as 40 Eridani A, which is the host star of the planet Vulcan in the Star Trek universe.} is stellar activity}
\author[0000-0002-5922-4469]{Abigail Burrows}
\affiliation{Department of Physics and Astronomy, Dartmouth College, Hanover, NH 03755, USA
}
\affil{\JPL}

\author[0000-0003-1312-9391]{Samuel Halverson}
\affil{\JPL}

\author[0000-0002-9337-0902]{Jared C. Siegel}
\affil{Department of Astrophysical Sciences, Princeton University, 4 Ivy Lane, Princeton, NJ 08540, USA}

\author[0000-0002-1743-3684]{Christian Gilbertson}
\affiliation{Center for Computing Research, Sandia National Laboratories, Albuquerque NM 87185 USA}
\affil{\PSUAA}
\affil{\PSUCEHW}

\author[0000-0002-4927-9925]{Jacob Luhn}
\affil{\UCI}

\author[0000-0002-0040-6815]{Jennifer Burt}
\affil{\JPL}

\author[0000-0003-4384-7220]{Chad F.\ Bender}
\affil{\UA}

\author[0000-0001-8127-5775]{Arpita Roy}
\affil{\STScI}
\affil{\JHU}

\author[0000-0002-4788-8858]{Ryan C. Terrien}
\affil{\Carleton}

\author[0009-0006-8612-0995]{Selma Vangstein}
\affil{\Carleton}

\author[0000-0001-9596-7983]{Suvrath Mahadevan}
\affil{\PSUAA}
\affil{\PSUCEHW}

\author[0000-0001-6160-5888]{Jason T.\ Wright}
\affil{\PSUAA}
\affil{\PSUCEHW}
\affil{\PSETI}

\author[0000-0003-0149-9678]{Paul Robertson}
\affil{\UCI}

\author[0000-0001-6545-639X]{Eric B. Ford}
\affil{\PSUAA}
\affil{\PSUCEHW}
\affil{\PSUICS}
\affil{\PSUCASt}

\author[0000-0001-7409-5688]{Guðmundur Stefánsson} 
\affil{NASA Sagan Fellow}
\affil{Department of Astrophysical Sciences, Princeton University, 4 Ivy Lane, Princeton, NJ 08540, USA}

\author[0000-0001-8720-5612]{Joe P.\ Ninan}
\affil{\PSUAA}
\affil{\PSUCEHW}

\author[0000-0002-6096-1749]{Cullen H.\ Blake}
\affil{\Penn}

\author[0000-0003-0241-8956]{Michael W.\ McElwain}
\affil{\GoddardESAL} 

\author[0000-0002-4046-987X]{Christian Schwab}
\affil{\Macquarie}

\author[0000-0001-5290-2952]{Jinglin Zhao}
\affil{\PSUAA}

\begin{abstract}
We revisit the long-studied radial velocity (RV) target {\target} using recent observations from the NASA-NSF \lq{}NEID\rq{} precision Doppler facility. Leveraging a suite of classical activity indicators, combined with line-by-line RV analyses, we demonstrate that the claimed 45-day signal previously identified as a planet candidate is most likely an activity-induced signal. Correlating the bulk (spectrally-averaged) RV with canonical line activity indicators confirms a multi-day \lq{}lag\rq{} between the observed activity indicator time series and the measured RV. When accounting for this lag, we show that much of the observed RV signal can be removed by a linear detrending of the data. Investigating activity at the line-by-line level, we find a depth-dependent correlation between individual line RVs and the bulk RVs, further indicative of periodic suppression of convective blueshift causing the observed RV variability, rather than an orbiting planet. We conclude that the combined evidence of the activity correlations and depth dependence is consistent with a radial velocity signature dominated by a rotationally-modulated activity signal at a period of $\sim$42 days. We hypothesize that this activity signature is due to a combination of spots and convective blueshift suppression. The tools applied in our analysis are broadly applicable to other stars, and could help paint a more comprehensive picture of the manifestations of stellar activity in future Doppler RV surveys.

\end{abstract}


\section{Introduction} \label{sec:intro}
To reach the precision necessary to detect temperate, Earth-mass extrasolar planets (exoplanets) around Sun-like stars using the radial velocity (RV) technique, the community must improve Doppler measurement precision significantly from the current state of the art ($\sim$0.5-1 {\ms}) to $\sim$10 {\cms} or better. Detecting and characterizing these exo-Earths is vital for future space-borne direct imaging missions, which will set the scientific priorities for the coming decade. With the latest generation of Doppler RV facilities such as NEID \citep{Halverson-2016}, EXPRES \citep{Jurgenson-2016}, ESPRESSO \citep{Pepe-2021}, and Maroon-X \citep{Seifahrt-2018} all demonstrating sub-m s$^{-1}$ RV precision over months-long timescales, uncorrected stellar activity is rapidly becoming the largest barrier to improving RV measurement capabilities.

To this end, the community has embarked on a wide range of recent studies aimed at using new techniques to diagnose and model periodic and quasi-periodic spectroscopic activity signatures in Sun-like stars, including the Sun \citep{deBeurs-2021, Zhao2023}. These explorations include line-by-line studies \citep{Siegel2022, Dumusque-2018, Cretignier_2020, AlMoulla2022, Siegel2022}, machine-learning-based approaches \citep[e.g.,][]{deBeurs-2021}, and novel techniques for detrending against classical activity indicators \citep{CollierCameron-2019}. A multitude of advancements have followed these studies, including the ability to isolate the velocities of individual, activity-correlated lines \citep{Cretignier_2020, AlMoulla2022}, and the discovery of a potential phase lag between the RV signals of active regions and line-asymmetry \citep{CollierCameron-2019}. These studies largely conclude that these metrics, i.e. RVs, line-morphology, and activity indices, may be different manifestations of a common latent activity process. Combining these advancements and techniques in one study might not only further prove that these measures are all related but also provide a deeper view into their influence on how we interpret a star's radial velocity signal.

Here we present a multi-pronged approach for identifying the source of periodic radial velocity signals, focusing on the long-studied target {\target}. {\target} is the subject of multiple investigations into the source of its velocity signature \citep{Ma2018, Diaz_2018}. \citet{Ma2018} concluded that an 8.47 Earth-mass planet orbits {\target} with a period of 42.4 days, very near to their claimed rotation period of the star (39-44.5 days, based on activity indicators), is the likely explanation for the observed RV signal. 

We leverage a suite of new or updated activity analyses to show that the observed signal is likely due to activity, rather than a planet. This hypothesis is supported by a wealth of activity indicators at the full spectrum and line-by-line level. Though the purpose of this paper is to demonstrate that {\target}’s observed RV variability is driven by activity, the outlined approach to reach this result could be widely applicable to similar targets. This sort of analysis could result in improved RV sensitivity to small planets and a better characterization of the underlying stellar activity for a wide range of targets.

The paper is organized as follows: Section~\ref{sec:hd26965} describes the stellar properties of {\target} and details of the previously proposed planet candidate. In Section~\ref{sec:neid_data}, we describe the NEID instrument, data products, and observations used in this study. Section~\ref{sec:methods} presents an overview of the various analysis pipelines applied to the NEID {\target} data products, and the computed quantities used in our line-by-line and depth analyses. In Section~\ref{sec:results}, we present an overview of our results comparing our computed line parameters (RV, depth, etc.) to a variety of activity metrics, including canonical spectral activity indicators as well as bulk integrated RVs.

\section{HD 26965}\label{sec:hd26965}

\subsection{Stellar properties}
{\target} is an ideal target for studying impacts of stellar activity given its inherent brightness ($V$ = 4.4), slow rotation period ($\sim$42 days), and generally low level of activity. The stellar parameters of {\target} are listed in Table~\ref{tab: stellParam}. Of note are {\target}’s moderate activity level of $\log R_{HK}^{\prime} = -4.99$ and low rotational velocity of $v \sin i =$ 1.23 km s$^{-1}$. These values are similar to the Sun, with ($\log R_{HK}^{\prime} = -4.91$ and $v \sin i = 1.6 \pm 0.3$  km s$^{-1}$). These collective properties make {\target} a tantalizing target for future space direct imaging missions and indeed, {\target} is listed as a high priority target for the Habitable Worlds Observatory \citep{Mamajek2023}.

\begin{table}[!htb]
\centering
    \caption{Reference stellar parameters of {\target} \citep{Diaz_2018, gaia, gaiadr3}.} 
    \begin{tabular}{c c c}
        \hline\hline
        Parameter & Value & Source\\
        \hline
        R.A. (J2000) & 04:15:16.32 & Gaia DR3 \\
        
        Decl. (J2000) & -07:39:10.34 & Gaia DR3\\
        
        $m_V$ & 4.43 & \citet{VizieR} \\
        
        \textit{B-V} & 0.82 & \citet{VizieR}\\
        
        Distance (pc) & $4.98 \pm 0.01$ & \citet{vanLeeuwen2007}\\
        
        \hline
        Spectral type & K0.5V & \citet{Gray2006}\\
        
        Mass ($M_\odot$) & $0.76 \pm 0.03$ & \citet{Diaz_2018}\\
        
        Age (Gyr) & $9.23 \pm 4.84$ & \citet{Diaz_2018}\\
        
        Luminosity ($L_\odot$) & 0.44 & \citet{anderson2012}\\
        
        $T_{\mathrm{eff}}$ (K) & $5151 \pm 55$ & \citet{Diaz_2018}\\
        
        [Fe/H] & $-0.29 \pm 0.12$ & \citet{Diaz_2018}\\
        
        log \textit{g} & $4.45 \pm 0.04$ & \citet{Diaz_2018}\\
        
        \textit{v} sin \textit{i} (km s$^{-1}$) & $1.23 \pm 0.28$ & \citet{Diaz_2018}\\
        
        log $R_{HK}^{'}$ & -4.99 & \citet{Jenkins2011}\\
        
        \hline
    \end{tabular}

    \label{tab: stellParam}
\end{table}

\subsection{Previous studies}\label{sec:prevStudies}

\citet{Diaz_2018} investigated the periodic {\target} bulk radial velocity signal using 1,111 RV measurements spanning 16 years from four instruments, the High Resolution Echelle Spectrograph (HIRES) at the Keck I telescope in Hawaii \citep{Vogt1994}, the Carnegie Planet Finder Spectrograph (PFS) at Las Campanas Observatory in Chile, the CHIRON high-resolution spectrometer \citep{Tokovinin2013} at the Cerro Tololo Interamerican Observatory (CTIO) in Chile, and the High Accuracy Radial Velocity Planet Searcher (HARPS) at the La Silla Observatory in Chile \citep{Mayor2003}. They found a best-fit Keplerian with a mass of $6.92\pm0.79$ $M_{\oplus}$ and a period of $42.364 \pm0.015$ days, but did not ultimately come to a definitive stance on the origins of the signal given their inability to rule out activity as a possible source. \citet{Diaz_2018} used canonical activity indicators as a proxy for magnetic activity and measured the periodicity of each indicator and its correlation with the RVs. From the HARPS data, they used the chromospheric Calcium II H\&K line index, $S_\mathrm{HK}$, H$\alpha$, the full-width half-max (FWHM) of the cross-correlation function (CCF), and the CCF bisector slope (BIS). They additionally used archival $S_\mathrm{HK}$ measurements from PFS and HIRES. They did not find statistically significant power in any of these indicators near their assumed stellar rotation period or 38 days, based primarily on archival photometric measurements, and thus found low correlation values with the radial velocities. However, when they considered $S_\mathrm{HK}$ measured from the original Mt.~Wilson data, they did find a strong signal near the observed radial velocity period (42 days). This is a key result of \citet{Diaz_2018} that casts uncertainty on the Keplerian nature of the observed Doppler signal.

\citet{Diaz_2018} directly tested the validity of a Keplerian signal by fitting a planet to the 16-year baseline of data and observing the stability of the fit over time. However, given the correlation between the Mt.~Wilson $S_\mathrm{HK}$ measurements and the velocity signal, \citet{Diaz_2018} concluded that more comprehensive modeling of stellar activity was needed to confirm a Keplerian source for the modulation.

\citet{Ma2018} conducted a similar analysis, but also included additional data from the Dharma Planet Survey (DPS) using the TOU high resolution optical spectrograph on Mt. Lemmon in Arizona. They also introduced an additional analysis on the invariance of their best-fit Keplerian to magnetically quiet or magnetically active periods to claim that the radial velocity signal of {\target} is likely driven by a planet of $8.47\pm0.47$ $M_{\oplus}$ with a period of $42.38 \pm0.01$ days.

Unlike \citet{Diaz_2018} and more similar to our findings, \citet{Ma2018} found that the $S_\mathrm{HK}$ index does show a clear modulation near the 42 day period. Despite the similar period of modulation, they find that the indicator only weakly correlates with the velocities.

Similar to \citet{Diaz_2018}, \citet{Ma2018} first test that their best-fit Keplerian is invariant in both period and amplitude across the timescale of their multiple data sources. Additionally, \citet{Ma2018} uses the Calcium HK index to identify two epochs: an ‘active magnetic phase’ and a ‘quiet magnetic phase.’ They find that in the active magnetic phase, $S_\mathrm{HK}$ is a factor of two larger than in the quiet magnetic phase. According to \citet{Lanza_2016}, solar RV variation and the level of chromospheric activity measured using $S_\mathrm{HK}$ are positively correlated. \citet{Ma2018} find that while the $S_\mathrm{HK}$ strength varies by a factor of two between the two phases, the best-fit Keplerian amplitude to the 42.38 day signal in the active phase is $1.7\pm 0.3$ m s$^{-1}$ and $1.8\pm 0.3$ m s$^{-1}$ in the quiet phase. They argue that invariance of the RV amplitude at the 42 day period supports the existence of the planet.

{\target} continues to be revisited. \citet{Rosenthal2021} combined over 30 years of RV surveys to investigate both existing and new exoplanetary signals within the legacy data. They deem HD26965 a false positive, with the signal categorized as a combination of long term activity and stellar rotation.

In a review of stellar and planetary signals investigated using EXPRES data, \citet{Zhao_2022} touch on HD26965 as a benchmark star for stellar activity. They show periodograms for a variety of different methods used to parse data of {\target} between ``cleaned'' and ``activity'' epochs. They find that, much like \citet{Diaz_2018}, they could not definitively determine whether the signal was Keplerian or not. They found that six of their methods of RV cleaning left behind RVs with no sign of the \citet{Ma2018} period, while eleven still did show signs of the period.

Most recently, \citet{Laliotis_2023} returned to the target as part of a larger survey of sun-like stars with host planets for potential direct imaging. They also argue that the signal is likely activity, as they find a significant H$\alpha$ signal at a period of about $43.5$ days, very close to the proposed \citet{Ma2018} signal though not quite at the 42 day period, further placing the planet hypothesis into question.

\section{NEID spectra} \label{sec:neid_data}
NEID \citep{Schwab-2016} is a highly stabilized, high resolution (R$\sim$115,000) precision RV spectrometer for the 3.5 m WIYN telescope. Since beginning science operations in Fall 2021, NEID has demonstrated $<$1 {\ms} performance on bright, quiet stars, including the Sun \citep{Lin-2022}. All NEID spectra of {\target} used in this study were collected through the NEID Earth Twin Survey \citep[NETS;][]{Gupta2021}. In total, our dataset included 63 spectra collected between October 16, 2021 and March 12, 2022. Raw 2D frames were reduced to 1D spectra using the standard NEID data extraction pipeline (version 1.2), which produces 1D spectra, RVs, cross-correlation functions (CCFs), and a suite of canonical line-index activity indicators for each stellar observation\footnote{https://neid.ipac.caltech.edu/docs/NEID-DRP/}. Our 63 observations have an average SNR of $\sim$420 per pixel at $\lambda =  550$ nm in the extracted spectra. Bulk velocities are computed using the CCF technique with a spectral mask \citep{Baranne1996, Pepe2002}. For this target, we use lines included in the ESPRESSO K2 spectral mask\footnote{available at https://www.eso.org/sci/software/pipelines/}, as this is closely matched to the target spectral type and is well-vetted for similar targets. For consistency, we also restrict our line-by-line RV analyses to the same lines in this mask to best compare results between the CCF-derived RVs (computed by the NEID data reduction pipeline) and our custom line-by-line pipeline. 

\begin{figure*}[htb!]
\centering
\includegraphics[width=0.8\textwidth]{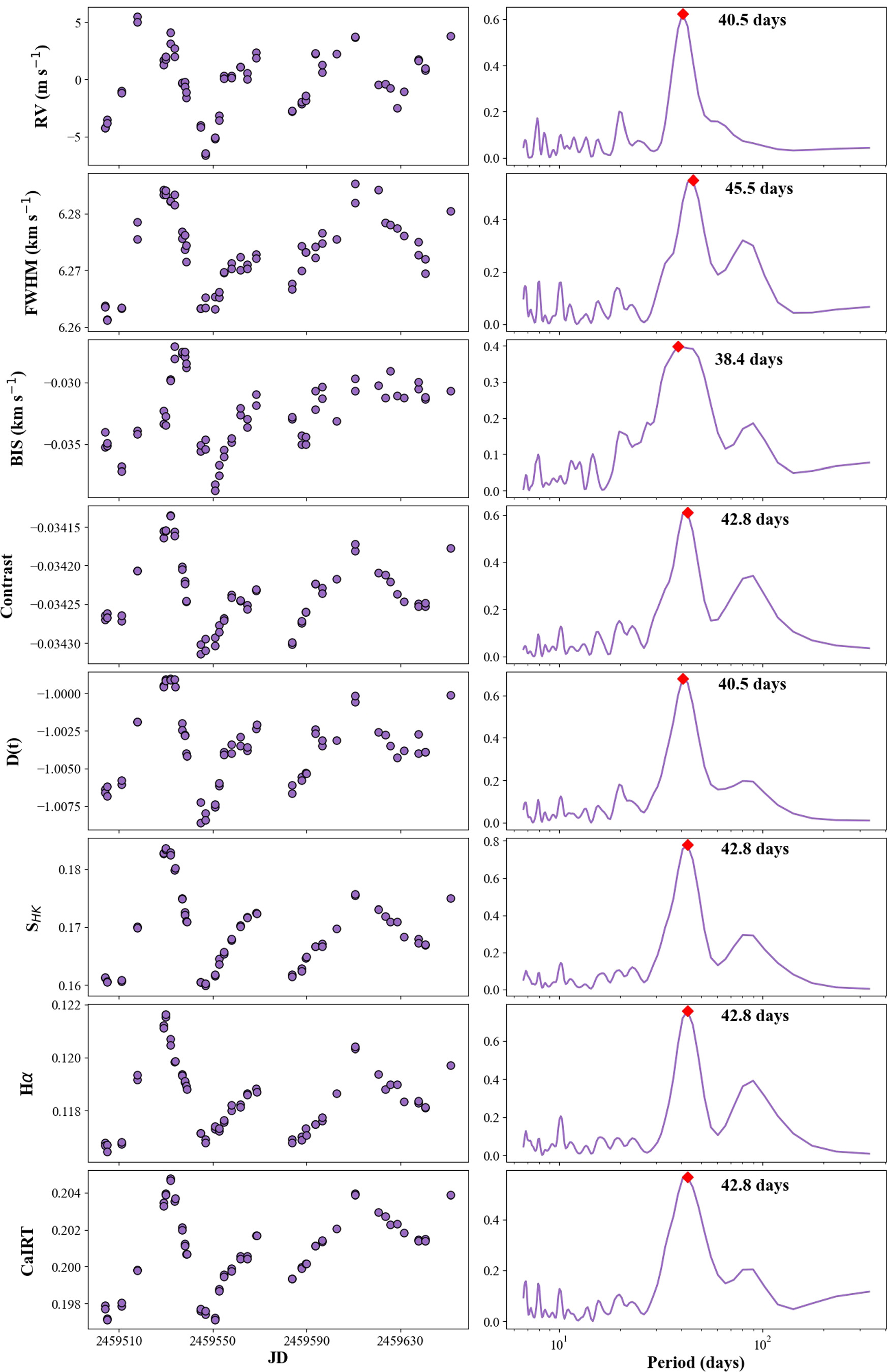}
\caption{\textit{Left}: NEID data of {\target}, showing velocities and activity indicator time series for spectra used in this study. Computed data products include velocities, CCF FWHM, BIS, contrast, $S_\mathrm{HK}$ index, H-$\alpha$ index, Ca IR triplet index, and depth metric \citep[see Section~\ref{subsec:depthMetrics}][]{Siegel2022}.  The red diamonds show the peak period in each periodogram. All data show clear rotational modulation at or near the 42 day period. \textit{Right}: Corresponding periodograms showing clear power at the stellar rotation period of $\sim$42 days.}
\label{fig:actIndPeriodograms}
\end{figure*}

The NEID data of {\target} used in this study are summarized in Figure \ref{fig:actIndPeriodograms}. The DRP computes bulk velocities, as well as a wealth of line  activity indicators derived from the 1D spectra. In addition to canonical line indices produced by the DRP, we separately compute the CCF full-width-half-maximum, (FWHM), contrast, and bisector inverse slope (BIS), as these have classically been used to diagnose activity. Figure \ref{fig:actIndPeriodograms} shows all of these data products for our NEID data of {\target}. All measurements show a strong signal at or near the 42-day stellar rotation period.

\section{Methods} \label{sec:methods}
Our investigation focuses on using NEID spectra and an expanded suite of analysis methods to reconsider the source of {\target}’s periodic radial velocities. The NEID spectra are measurably higher cadence and higher signal-to-noise (SNR$\sim$400) than previous studies, allowing for a deeper study of rotationally-modulated activity signatures in {\target} at the individual spectral line level. 

To address the weak correlations found by both \citet{Diaz_2018} and \citet{Ma2018} between the various activity indices and the integrated radial velocities, we introduce the possibility of a phase lag between the indicators and the RVs. \citet{CollierCameron-2019} present evidence for a phase lag between the effect of activity in line morphology and the corresponding RV as a result of modulating spots or plages in solar data. We hypothesize that this may cause the velocity effects of activity to appear weakly correlated with activity indicators, even if they are modulated at identical periods.

Finally, we calculate the phase folded RV signal for magnetically active and quiet lines, and find the best-fit Keplerian for the resulting time series. We hypothesize that a velocity signal dominated by activity will find best-fit Keplerians with different amplitudes for the populations of quiet and active lines. We argue for this approach given that the stability of the integrated RV signal may be dominated by long-term activity. 

\subsection{Line-by-line pipeline description}
We begin our activity analysis by employing a line-by-line (LBL) measurement pipeline to explore previously noted manifestations of rotationally-modulated activity. The LBL approach computes an integrated velocity signal by calculating the individual radial velocities of spectral lines across a spectrum, and combining them as weighted by the photon-limited error on each line velocity. This LBL approach has been demonstrated to be very powerful for characterizing activity signals in HARPS data \citep{Dumusque-2018, Cretignier_2020, Siegel2022}. Our line library, derived from the ESPRESSO K2 mask, contains 6592 lines from 3700~{\AA} to 7800~\AA. For each line in this mask, we measure the velocity, line depth, and line symmetry.

\subsubsection{Data pre-processing} \label{sec:conditioning}
To continuum normalize the reduced 1D spectra, we employ two different blaze removal methods, one for the RV calculation and another for the line morphology. The NEID pipeline natively produces an order-by-order model of the grating blaze profile for each observation, derived using a combination of broadband lamp sources (see NEID DRP documentation). Dividing each stellar spectral order by this empirically-modeled blaze greatly flattens the stellar spectra. For the RVs computed in our LBL pipeline, we find that dividing by the DRP-produced blaze function is sufficient for measuring precise and repeatable individual line RVs, and is in fact preferable to additional steps such as spline removal as it preserves the most natural information regarding the weighting of each line and is computationally efficient.

To measure the most accurate line depths, however, we find that additional continuum normalization is preferable for identifying key line boundary points such as local minima and maxima. We use a modified version of the Rassine package \citep{Rassine2020}, which makes use of both a convex-hull and rolling alpha method, to flatten the spectrum prior to making depth measurements. For our use, this method yields more repeatable and accurate line depth measurements relative to the local continuum.

\subsubsection{Template creation}
\label{subsec:templateCreation}
In order to measure velocity shifts of individual spectral features, we compare each line to a high signal-to-noise template spectrum. To generate this template, we shift each wavelength solution to the stellar rest frame and co-add all 63 individual stellar spectra (shifted to the stellar rest frame).

Each NEID spectrum is measured in the reference frame of the observatory, and therefore must be corrected for the changing Earth-Sun velocity vector to produce RVs in the stellar frame. To generate a high SNR spectral template, we first shift each file's wavelength solution to the stellar rest frame by subtracting the barycentric velocity, systemic velocity (\lq{}\texttt{QRV}\rq{} in the NEID headers), and the velocity of each file as derived from the DRP. This ensures all spectra being coadded are in as similar of a velocity frame as possible (limited at the accuracy of the derived RVs).

Finally, to ensure that each observation shares the same wavelength grid, we linearly interpolate (using \texttt{scipy} \texttt{interp1d}) the flux of each file onto a single wavelength solution. For this solution, we select the wavelength grid of the first file, shifted to the appropriate rest-frame velocity by removing systemic and barycentric velocity components, as the \lq{}template\rq{} solution for all files.

Once each file is properly blaze-corrected, shifted, and interpolated, we sum the interpolated flux arrays to create the high SNR template against which we can measure the velocities of individual spectral features.

\subsubsection{Measuring individual spectral line RVs}
\label{subsec:lineRVs}
 To compute line-by-line (LBL) RVs, we largely follow the methodology outlined in \cite{Dumusque-2018} and \cite{Cretignier_2020}. For completeness, we provide a brief summary of the methodology here. Each line $i$, at wavelength $\lambda_i$, in the stellar spectrum has its own independent radial velocity, $RV_i$. For each line, we focus on a 16 pixel ($\sim$9 k{\ms}) bin around line center in the template ($S_{\mathrm{temp},i}(\lambda)$). This is an approximately 0.015 nm-wide bin at the NEID pixel sampling at 480 nm, and is comparable in width to the region of the NEID DRP CCF that is used to fit the CCF RVs. We then compare each line chunk in the template to each observation, denoted by $S_{j,i}(\lambda)$. Line-by-line velocities are measured by simultaneously fitting the amplitude ($A$) and wavelength offset ($\delta\lambda$) between parameters $S_{\mathrm{temp,i}}(\lambda)$ and $S_{j,i}(\lambda)$. We assume each spectral chunk in a given observation, $S_{j,i}(\lambda)$, can be parameterized as:

\begin{equation}
    S_{j,i}(\lambda) = A\left[S_{\mathrm{temp,i}}(\lambda) + \frac{\partial S_{\mathrm{temp,i}}(\lambda)}{\partial \lambda}\delta\lambda\right]
\label{eq:rv_fit}
\end{equation}

Before fitting $A$ and $A\delta\lambda$, we apply a Doppler shift to the template to match a given observation’s rest frame velocity. This shift corrects for both the systemic velocity and the observation's barycentric velocity. Next, we linearly interpolate and resample the template spectrum to match the observation's wavelength solution. Once resampled to the observation's wavelength solution, we then apply each spectrum's model blaze function, derived for each NEID L1 spectrum using a combination of flats, to the interpolated template such that the overall intensity profile across each echelle order matches the observation. In all cases, we prefer to manipulate the flux values in the high SNR template rather than the individual spectra to avoid added interpolation errors.

Once the template is properly re-weighted, we fit for $A$ and $A\delta\lambda$. We use a non-linear least squares regression to fit both parameters and derive uncertainties using the computed covariance matrix. Even with the local blaze model scaled to each observation, fitting for both $A$ and $A\delta\lambda$ simultaneously is necessary to recover reliable RVs for all lines, as the remaining small-scale continuum offsets between the template and observation otherwise add noise to the fitted velocities. We then compute the Doppler shift, $\mathrm{RV}_{i,j}$, using the fitted values of $A$ and $A\delta\lambda$:

\begin{equation}
    RV_{i,j} = \frac{c}{\lambda} \frac{A\delta\lambda}{A}
\end{equation}
with corresponding error
\begin{equation}
    \sigma_{\mathrm{RV}_{i,j}} = \frac{c}{\lambda} \sqrt{\left[\frac{1}{A}\right]^2 \sigma_{A\delta\lambda}^2 + \left[-\frac{A\delta\lambda}{A^2}\right]^2\sigma_{A}^2}
\end{equation}.

Given the high SNR of the NEID spectra, our individual line RV uncertainties reach at their best an error of 3.3 ms$^{-1}$, with a median time-series-average error of 37 ms$^{-1}$ per line. The measured time series RMS of any given line is 38 ms$^{-1}$ (median), in reasonable agreement with the individual line white-noise error bars.

\subsubsection{Sigma clipping individual lines}
\label{subsec:clipRVs}

Once individual line RV time series are computed, we then perform a series of sigma cuts to remove highly discrepant lines that may bias the overall velocity signal. Following the filtering prescription of \citet{Dumusque-2020}:
\begin{itemize}
\item[--] We perform a 6-$\sigma$ clip on all RVs in a given observation.
\item[--] We perform a 6-$\sigma$ clip on the residual between each line velocity time series and a fitted second order polynomial.
\item[--] We reject all lines for which more than 1$\%$ of the observations are cut, which in this case is one observation or more.
\item[--] We remove the $0.3\%$ of lines with the highest overall time-series RMS values, in this case approximately 20 lines.
\item[--] We finally remove all lines where the ratio between the scatter in the RV time series and the time series' median error is greater than two to further shield against other potentially contaminated lines.
\end{itemize}

Following these cuts, we are left with 6400 of the original 6592 lines (97\%) in the K2 mask.

\subsubsection{Measuring bulk RV from spectral line RVs}
\label{subsec:bulkRVs}

Though the individual lines are interesting because of the detail with which we can divide the signal into its components, their lack of integrated information content can make it difficult to draw confident inferences. We can boost the overall RV signal-to-noise by combining many lines at once into an integrated RV measurement. The integrated RV is calculated by taking the sum of individual line RVs weighted by their error:
\begin{equation}
    \label{eqn:rv_weighted_avg}
    RV_j = \frac{\Sigma_i \left[\frac{1}{\sigma^2_{RV_{i,j}}}RV_{i,j}\right]}{\Sigma_i \left[\frac{1}{\sigma^2_{RV_{i,j}}}\right]}
\end{equation}

\begin{figure*}
\centering
\includegraphics[width=0.7\textwidth]{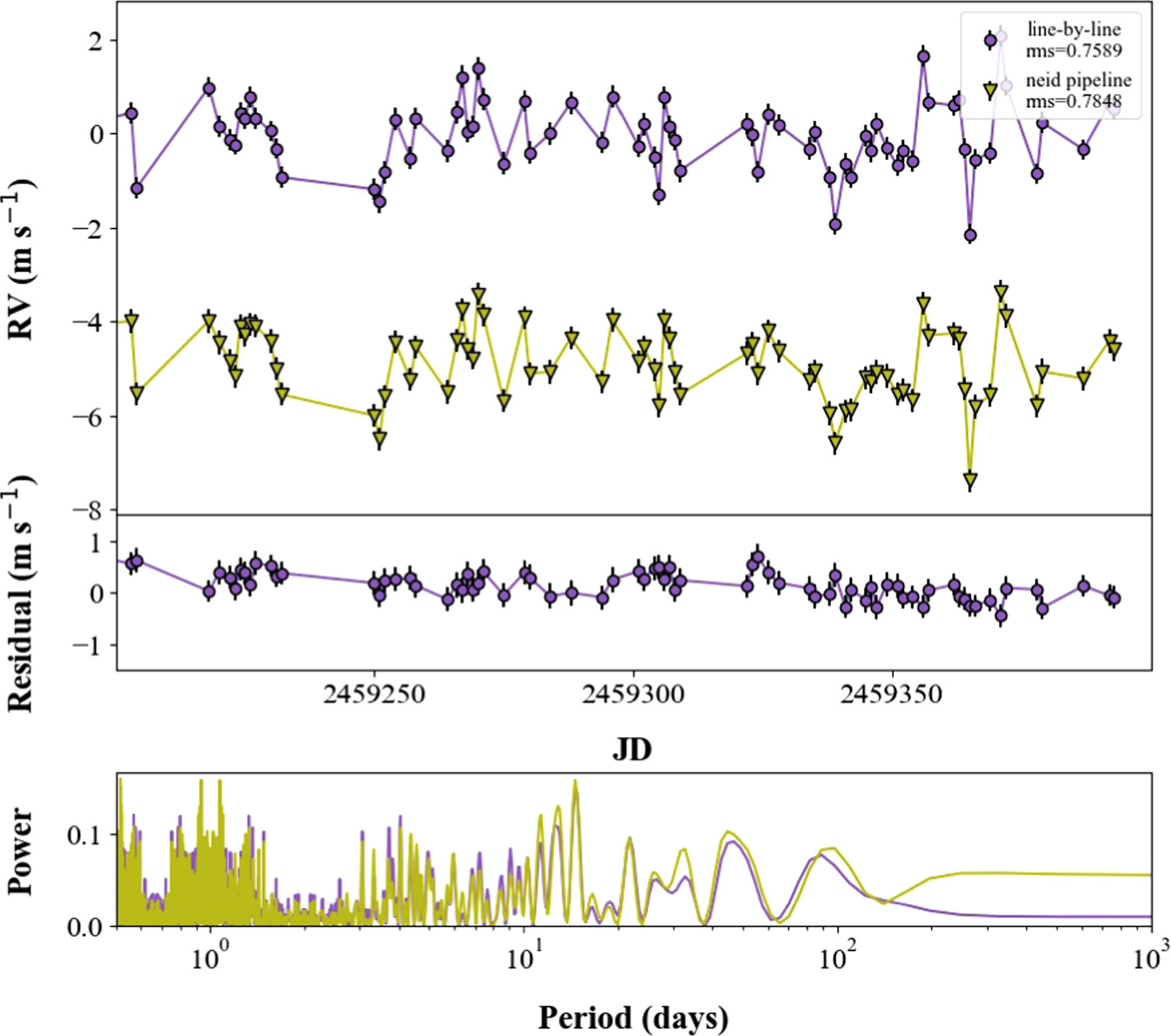}
\caption{Demonstration of our line-by-line (LBL) on NEID solar spectra, compared to the standard NEID data reduction pipeline (DRP). To compute the LBL RVs, we constructed a high SNR solar template using NEID spectra from an exemplary “good weather” day, which includes 192 spectra. \textit{Top}: Solar bulk RVs calculated using LBL pipeline (purple) as compared to the NEID DRP (yellow) using the same spectral lines (offset added for clarity). Data were manually selected for days with no measurable clouds. Our LBL pipeline produces nearly identical results for the full time series. \textit{Middle}: Residual time series shows little structure and are broadly consistent with the photon errors.
\textit{Bottom}: Periodograms of NEID pipeline and line-by-line RVs, highlighting the similar temporal structure.}

\label{fig:solarIntVels}
\end{figure*}

\begin{figure*}
\centering\includegraphics[width=0.7\textwidth]{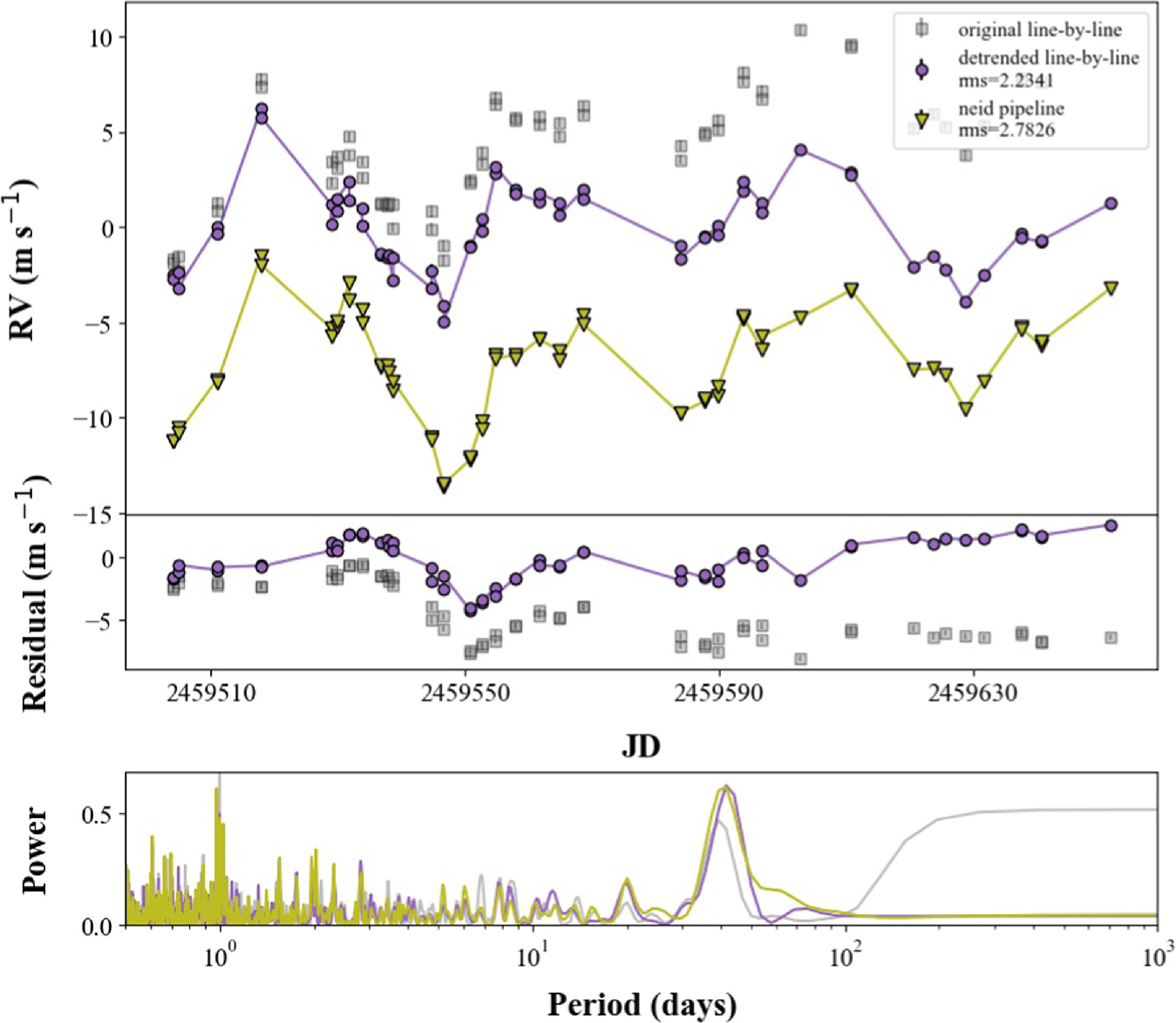}
\caption{\textit{Top}: HD26965 bulk RVs calculated through our line-by-line (LBL) pipeline (grey) with linear trend removed (purple) as compared to the NEID pipeline RVs (yellow, offset added for clarity). \textit{Middle}: Residual time series after linearly detrending the LBL RVs (purple) retains approximately 2.5 ms$^{-1}$ modulation in the first half of the time series. \textit{Bottom}: Lomb-Scargle (LS) periodograms of all three RV time series. All show same peak period of around 42 days. Linearly detrending LBL RVs (purple) removes the long period power seen in the LS periodogram, and results in a better agreement between the NEID pipeline RVs and the integrated LBL RVs. Possible causes for the differences between the two RVs are detailed in Section~\ref{subsec:bulkRVs}. As we are primarily interested in signals at or near the 42 day period, we remove this trend from the LBL bulk RV for the remainder of the analysis.}
\label{fig:hd26965IntVels}
\end{figure*}

Using this technique, we can compute a line-by-line-derived equivalent bulk RV measurement to compare against the CCF-derived RVs. To verify the performance of our pipeline, we apply our LBL pipeline to both publicly available NEID solar data  (as a methodological test) and data of {\target}. Figures \ref{fig:solarIntVels} and \ref{fig:hd26965IntVels} compare our line-by-line RVs to those derived using the standard NEID pipeline using the same set of spectral lines (ignoring lines that were filtered in the previous step). As we also are restricting our LBL analyses to these mask line wavelengths, we expect our RVs to be highly correlated. For both HD26965 and the sun, we find generally consistent agreement between the  CCF-based and LBL RVs. {\target} shows measurable residuals between the NEID pipeline and LBL RVs, though the signal periodicities are similar (Figure~\ref{fig:hd26965IntVels}). 

We have multiple hypotheses for the discrepancies in the RVs of {\target}. First, the CCF RVs do not use the \lq{}natural\rq{} line error weighting scale used by the LBL pipeline, instead fixing the relative contributions of each spectral order. This means that the LBL RVs may show slightly stronger signals from lines that are forcibly down-weighted by the CCFs and vice versa. Another potential difference arises from data conditioning. The LBL RVs are naturally computed on blaze-corrected spectra, while the CCF RVs are computed on the \lq{}blazed\rq{} spectra. This should not be significant, as the original pixel flux uncertainties are still tracked when computing the LBL RVs, but may produce slightly different results. This effect may be less for the solar spectra (which is generally consistent with Figure~\ref{fig:solarIntVels}, where we do not see a similar linear trend), where the relative motion of the stellar spectrum to the blaze function is small. We also noted several individual spectral order RVs measured in the NEID pipeline that were systematically significantly different from the bulk RVs, which may have biased the spectrally-averaged measurements in a systematic way.

In either case, Figures \ref{fig:solarIntVels} and \ref{fig:hd26965IntVels} show the resulting periodograms of both the CCF and line-by-line RVs. Importantly for this paper, the structure of the RV oscillations found in the CCF RVs are recovered by the line-by-line RVs. As such, we focus the remainder of our analysis on the LBL RVs alone.

\subsection{Computing Line Parameters}
\label{sec:lineParams}
To maximize sensitivity to potential activity signals in the LBL RVs, we select and group lines by different parameters, then investigate the integrated RV signals of these groups of lines. One natural way to group lines is by normalized line depth. This is particularly useful when considering activity, as line depth may be inversely correlated with strength of magnetic activity signal in the line's RV \citep{Cretignier_2020}. Below we present our method for calculating a similar suite of line parameters that lead to our measurement of depth.

\subsubsection{Line depth}
\label{subsec:edgePoints}
To measure the individual line depths, we broadly follow the prescription of \cite{Cretignier_2020}. For completeness, we summarize the general approach here. We define line depth as the maximum flux difference between the local minima and maxima of a line in a continuum-normalized spectrum. First, we search for the two local maxima points on the blue and red wings of the line. In order to reduce contaminating blends, we only search for maxima points within a 40 pixel bin around line center, a wider bin than used for computing the individual line RVs as the maxima and minima points of interest may be located further out from line center. For the NEID instrument, this is generally a 0.04 nm bin at 480 nm. In all cases, we consider the blue and red maxima closest to line center. In cases where the algorithm misses a local maximum, we force a second maximum to avoid unnaturally shrinking the line depth. The artificial second local maxima is the point on the opposite wing with the same velocity shift from empirically-measured line-center as the algorithm-discovered maxima. We then measure the flux difference between the local minima and each local maxima. We define the line depth as the maximum of these two flux measurements.

Using this definition of depth, we find a reliable depth measurement for all of the lines defined in our binary spectral mask. For all LBL analyses, we exclude lines with a normalized depth of less than 0.05 to reduce the likelihood of measuring RVs on stretches of continuum. After our depth cut, we are left with 5401 lines out of the original 6592.

\subsection{Line symmetry}
\label{subsec:lineSymm}
Careful curating of lines is an important step in mitigating against spurious RV. Though all lines have a natural asymmetry due to the stellar convective blueshift, blended lines will show particularly strong asymmetry. This asymmetry can be measured in the normalized spectra, and may be indicative of unresolved line blends that could add systematic noise to individual line RV measurements. We calculate our symmetry parameters on both the co-added template spectrum as well as on the time series of individual spectra to ensure that each line meets the symmetry cut across the entire time-series. Below we describe our procedure to diagnose line asymmetry, following the general outline presented in \cite{Cretignier_2020}:

\subsubsection{Symmetry selection}
\label{subsec:depthMetrics}
\begin{table*}[!htb]
\centering
    \caption{Parameters and values used to determine symmetric lines. Jerk distance and mass centre values were adapted from \citet{Cretignier_2020} in order to correct for systematic morphological differences observed between lines that are positively and negatively correlated with bulk-RV.} 
\begin{tabular}{c c c c }
        \hline\hline
        Parameter & Cut & Symmetric Value & Source\\
        \hline
        Continuum difference & < 0.3 & 0 & \cite{Cretignier_2020} \\
        Continuum average & > 0.8, < 1.2  & 1 & \cite{Cretignier_2020} \\
        Small window & > 0.004 & -- & \cite{Cretignier_2020}\\
        Jerk distance & < 0.15 & 0 & optimized for NEID\\
        Mass centre (derivative) & < 0.15 & 0 & optimized for NEID\\
        \hline
        
    \end{tabular}

    \label{tab:symmSelection}
\end{table*}

\begin{figure}[htb!]
\centering
\includegraphics[width=\columnwidth]{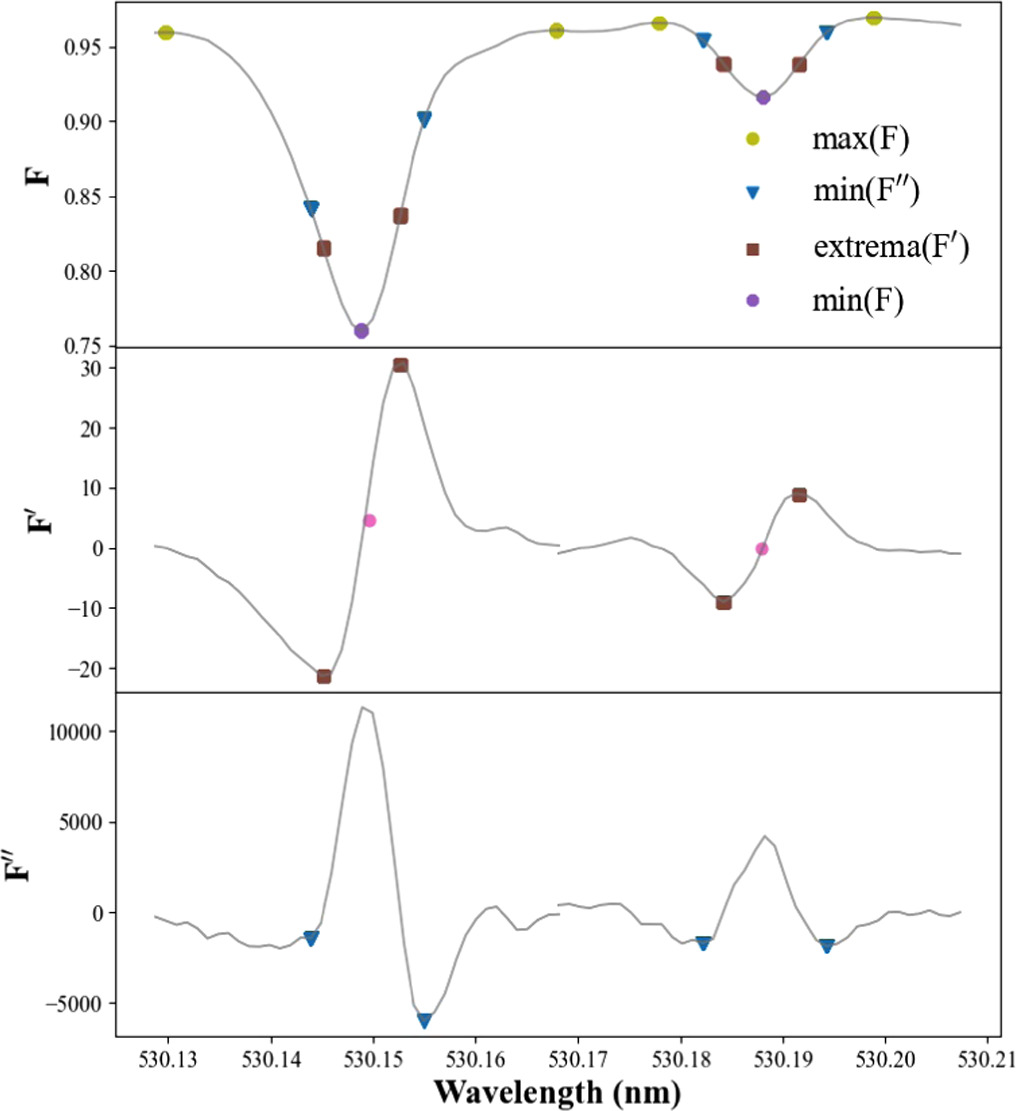}
\caption{Critical points for computing diagnostic symmetry metrics are plotted across the spectrum (F), its first derivative (F$^{\prime}$), and its second derivative (F$^{{\prime}{\prime}}$) in: yellow for local maxima (continuum average, continuum difference); blue for negative minima of the second derivative (small window, jerk distance); brown for extrema of the first derivative (mass centre); purple for the local minimum (depth). In this 0.8 nm portion of the spectra, the left-hand line is marked as asymmetric and the right-hand line as symmetric.}
\label{fig:lineSymm}
\end{figure}

We calculate line symmetry based on the five criteria listed in Table~\ref{tab:symmSelection}. When calculating normalized line depth, we first find the maxima on either side of the line using a combination of criteria derived from the flux values and the first and second derivatives. These points allow for the measurement of the first two constraints for symmetry: continuum difference and continuum average. Continuum difference is defined as difference in flux between the two edge maxima, while continuum average is the arithmetic mean. Figure~\ref{fig:lineSymm} shows an example of the points used to calculate each of these parameters. For non-continuous regions of the spectra where one local maximum is not identified (e.g. due to bad detector columns or hot pixels), we simply reflect the identified maximum about the line center to define the opposite edge of the feature (using the same velocity offset from the correctly identified maximum). We do this for both the normalized spectrum, as well as its derivative maxima/minima (see Figure~\ref{fig:lineSymm}).

Next, we calculate the mass centre of the line by taking a weighted average of the flux within the line between the two extrema of the first derivative (two-nearest inflection points). A perfectly symmetric line will find that the centre of mass of these two points lies directly at the wavelength of the minima, or $(\lambda_{min}, 0)$. We normalize the mass-centre derivative by dividing the normalized flux value of the centre of mass by the difference between the flux values of the inflection points.

Finally, we use the jerk distance to diagnose asymmetry of the line center with respect to the small window edges. The jerk distance is a measure of the difference between the depth calculated from the left bound of the small window and the right bound, normalized by the actual line-depth.

In determining criteria for line inclusion (Table~\ref{tab:symmSelection}), we follow the prescription of \citet{Cretignier_2020}. First, we only include lines with a jerk distance of $|jd| < 0.25$, and a mass-centre derivative of $|mcd| < 0.2$. In order to confirm that our criteria listed in Table~\ref{tab:symmSelection} are preferentially filtering out blended lines, we track both the total number of lines that pass each filter step, as well as the ratio between the number of lines that show a positive correlation with the bulk (averaged) RV and those that have a negative correlation (assumed to be an unresolved blend).

To ensure our symmetry cuts are indeed reliably removing  blends without providing a bias against lines with a truly negatively correlated RV signature, we calculate the median value of each of our symmetry parameters after the initial symmetry cut for both the negatively and positively correlated lines. If there are systematic morphological differences between these two populations, we expect that asymmetric lines likely dominate the negatively correlated lines. We find that, on average, the mass centre derivative is a factor of two larger and the jerk distance is a factor of 1/3 larger in the negatively-correlated lines than in the positively-correlated ones. When we further restrict the jerk distance to a more stringent value (0.15), the mass centre derivatives in both populations of lines falls under $ < |.15|$. This does not remove all the negatively-correlated lines from our analysis, but does significantly reduce the relative fraction of them and there are no longer significant morphological differences between the negatively and positively correlated lines, resulting in 1146 lines left at this stage.

\section{Results} \label{sec:results}
\begin{figure}
\centering
\includegraphics[width=\columnwidth]{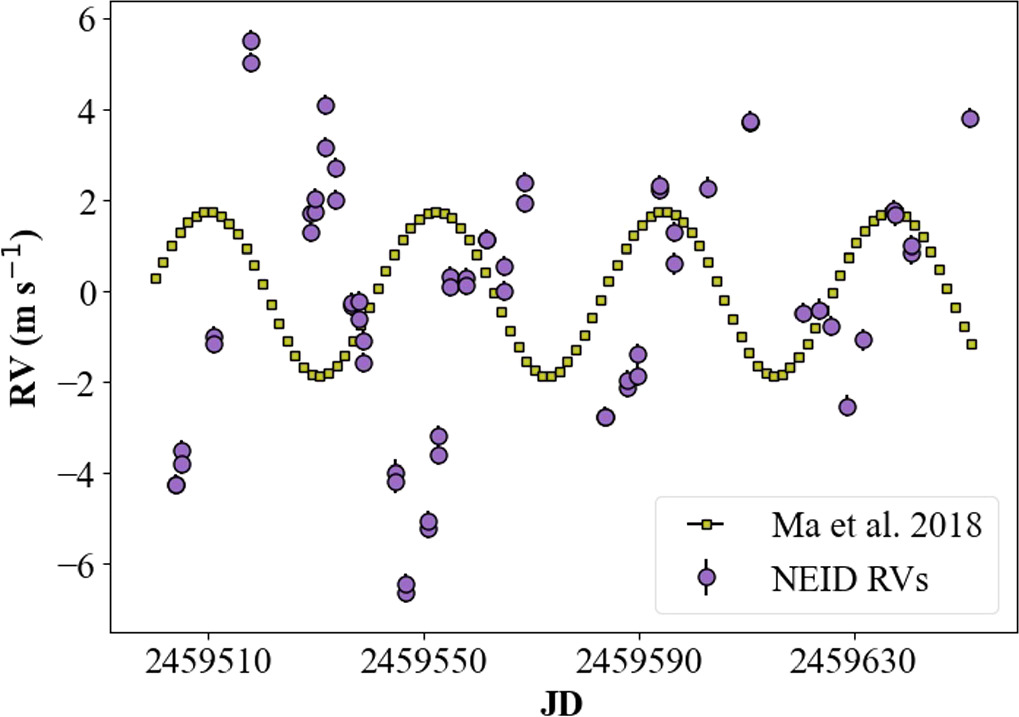}
\caption{NEID bulk RVs of {\target} (purple) overlayed on the best-fit Keplerian solution from \citet{Ma2018}. Within the errors of the \citet{Ma2018} (propagated to the NEID measurement epochs), the proposed Keplerian is significantly out of phase with the dominant 42 day signal in the NEID data.}
\label{fig:MaModel}
\end{figure}

With our arsenal of line and activity metrics in hand, we begin investigating the information embedded in individual line RVs. We do so using two main methods: 1) by correlating individual line metrics with independent activity metrics, and 2) by correlating individual line metrics with the bulk RV computed using all lines in the K2 mask. As an initial check of the robustness of the previous planet model, we first examine the phase of the observed NEID RVs relative to the original planet model posed in \cite{Ma2018}. Figure~\ref{fig:MaModel} shows the previous best-fit planet model is significantly out of phase (approximately 30-40\% of the expected period) with the observed 42 day signal in the NEID bulk RVs, casting doubt on the coherence of 42 day signal over long time baselines. This broadly supports an activity-driven RV hypothesis.

\subsection{Bulk RV Analysis}

We begin by investigating the correlation between our line-by-line-derived bulk-RVs and various classical activity indicators in the NEID data. Importantly, we explore the correlation between bulk RV and various activity indicators with and without accounting for a relative phase shift. This step was not explicitly done in \cite{Ma2018}, and appears to be key for identifying activity-induced RV signals in this particular target. In the following sections, we quantify the relative power at the proposed planet period, showing that it decreases significantly when including an added phase lag between various activity indicators and the bulk RVs.

\subsubsection{Bulk RV and $S_\mathrm{HK}$ Index correlation}\label{sec:bulkCorrs}
We begin our activity detrending analysis by exploring the linear correlation between Ca II H\&K and the bulk RVs. As shown in Figure~\ref{fig:actIndPeriodograms}, {\target}'s RV signal and Ca II H\&K time series both show clear rotational modulation at a similar period of $\sim$42 days. However, the linear correlation coefficient between the two time series is relatively low (Pearson's $R=0.38$). This indicates the signals are measurably out of phase over the several stellar rotations present in the NEID data.

\subsubsection{Other activity indicators}\label{sec:otherActInds}
We also investigated using other line indices, motivated by the discussion of the complexities of stellar activity signatures in \citet{Diaz_2018}. The NEID pipeline automatically calculates several other line index measurements, including H$\alpha$ and the Ca IR triplet (Ca IRT). We find that H$\alpha$ correlates nearly perfectly with $S_\mathrm{HK}$. We also find the Ca IRT index similarly strongly correlates with the $S_\mathrm{HK}$ index (as expected). A summary of each time series is shown in the left panel of Figure~\ref{fig:actIndPeriodograms}.

Similarly to the RV and $S_\mathrm{HK}$ correlation, linear correlations between the RVs and the Ca IRT and H$\alpha$ indices yield correlations of $R\sim0.36$. We additionally consider the CCF FWHM, contrast, and BIS, which nominally trace the changing shapes of spectral lines due to the rotation and evolution of active regions like spots or plages \citep{Costes_2021}. We measure the FWHM by directly fitting the CCF with a Gaussian function. We compute the bisector inverse slope by calculating the difference between the average mid-point velocity of the 10-40th percentile and the 60-90th percentile of the CCF line core.

As shown in the right panels of Figure~\ref{fig:actIndPeriodograms}, the dominant periodicity in the CCF FWHM and CCF BIS are also broadly consistent with the radial velocities at 44 and 39 days, respectively.%

\begin{figure*}[htb!]
\centering

\includegraphics[width=0.9\textwidth]{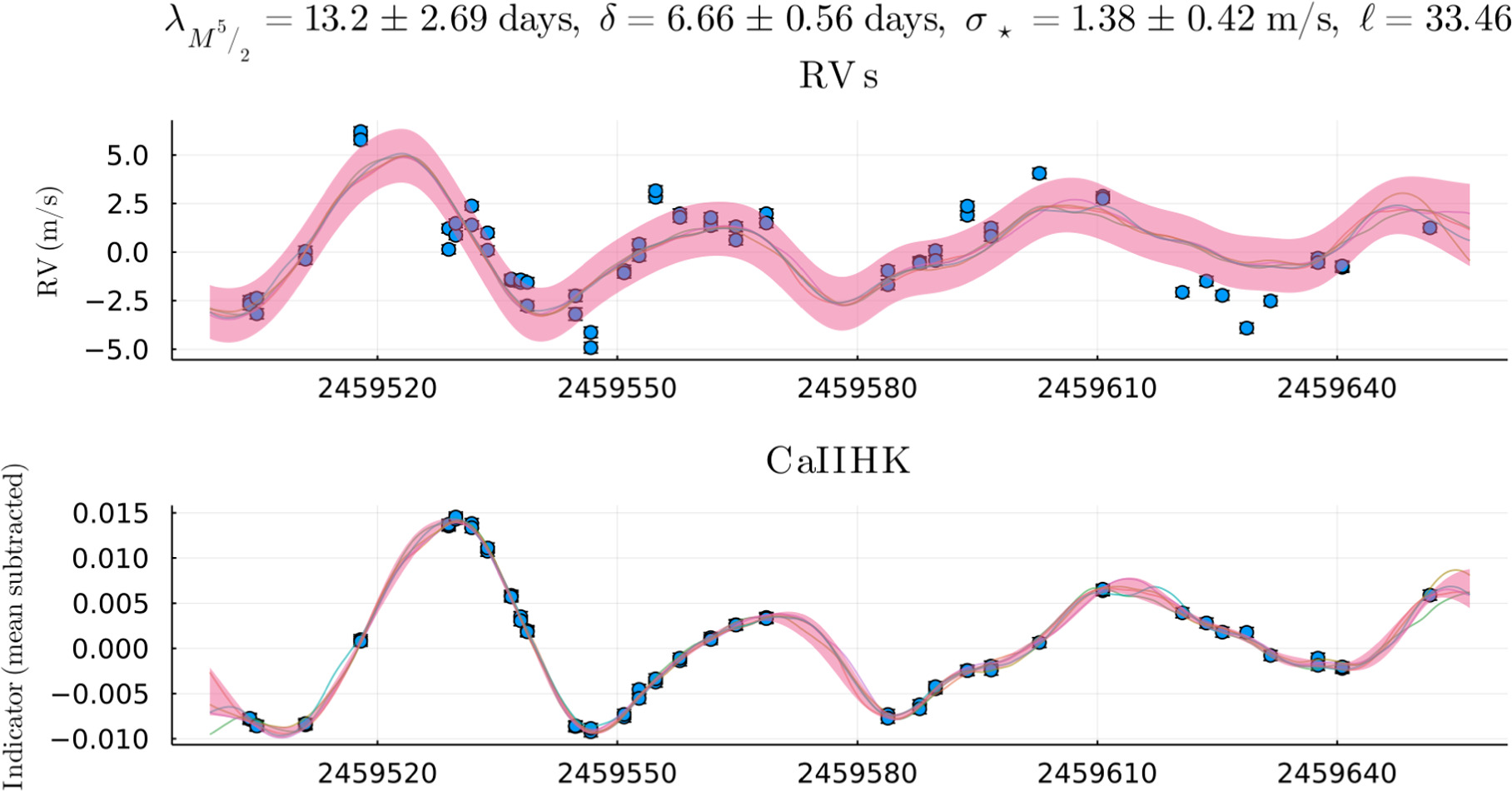}
\caption{Example GP-modeled phase lag between $S_\mathrm{HK}$ activity indicator and line-by-line bulk RV time series with an additional term included to account for smaller-time scale RV jitter. Our model detects a phase lag of $\sim$6.7 days between the $S_\mathrm{HK}$ index and the bulk RVs (with the RVs following behind the line indicator). Thin solid lines show a set of random draws from the model posterior distributions for clarity.}
\label{fig:gprvSinFitsCaHK}
\end{figure*}

\subsubsection{Calculating phase offset using Gaussian Processes}\label{sec:phase}
A phase offset between RV and magnetic activity is a known phenomena in spectra of Sun-like stars: an RV signal induced by a plage or a dark spot could induce their strongest RV signal out of phase from the magnetic activity strength \citep{CollierCameron-2019}. Without investigating a phase lag, activity indicators may appear poorly-correlated with even activity-dominated RV signals (as demonstrated in the previous section).

Using a shared latent Gaussian process (GP) model, $X(t)$, and an additional white-noise RV jitter term to model the two time series allows us to determine the relationship between the two time series without assuming that the underlying signal is of a fixed functional form. We use the GP model to independently derive the relative phase between the RVs and each activity indicator. We used \texttt{GPLinearODEMaker.jl} \citep[\texttt{GLOM};][]{gilbertson2020} to fit a single GP with a Mat\'ern $^5/_2$ kernel with a lag hyperparameter (see Equation~\eqref{eq:m52kf}) (and the RV jitter term) to the RVs and an indicator simultaneously. The Mat\'ern $^5/_2$ kernel is

\begin{equation}
 k_{M^5/_2}(t,t') \propto \left(1 + \Delta t + \dfrac{\Delta t^2}{3}\right) \ e^{-\Delta t}
 \label{eq:m52kf}
\end{equation}

\noindent where $\Delta t = \sqrt{5} \ |t-t'\pm\delta| / \lambda_{M^5/_2}$, $t-t'$ is the difference between the two observation times, $\delta$ is the lag hyperparameter describing the delay between the RVs and indicator ($\delta=0$ when $k_{M^5/_2}$ is being evaluated between two RVs or two indicator measurements and is added or subtracted depending on whether $k_{M^5/_2}$ is being evaluated between an RV and the indicator or vice versa) and $\lambda_{M^5/_2}$ is the timescale of local variations.
The RV and indicators are modeled as

\begin{equation}
 \text{RV}(t) = a_{RV}X(t) + \epsilon_{\sigma_\star} + \epsilon_{RV}
 \label{eq:model1}
\end{equation}

\begin{equation}
 \text{I}(t) = a_{I}X(t) + \epsilon_{I}
 \label{eq:model2}
\end{equation}

\noindent where $a$ are hyperparameters that control the relative amplitude of the GP components, $X(t)$ is the latent GP that is shared by the RVs and indicators, and $\epsilon_{\mathrm{RV}}$ and $\epsilon_{I}$ are measurement uncertainties and RV jitter terms. 
In essence, we are assuming that the RVs and indicator have the same, though unknown, shape and the only differences come from a time delay, differing amplitudes, and some extra white noise in the RVs. Figure~\ref{fig:gprvSinFitsCaHK} shows the resulting \texttt{GLOM} model for the $S_\mathrm{HK}$index, and Table~\ref{phaseShifts} presents the best-fit phase offsets for all activity indicators. We find a stable phase lag between approximately 4.65 and 6.67 days, which is approximately 11-15$\%$ of the rotation period.

\subsubsection{Calculating phase offset via interpolation}

To augment the GP-derived lags, we also use a simple interpolative model to separately measure the phase delay between signals. For each activity indicator, we measure the correlation coefficient at a variety of offsets and fit the cross-correlation function peak to estimate the lag. We compute correlations for offsets from 0 to 15 days, with steps of 0.5 days, using linear interpolation to fill in gaps in the data. To empirically estimate the error on our best-fit lag value, we shuffled the indices and the RVs within their 1-$\sigma$ errors and recomputed the nominal phase shift 1000 times. We fit a Gaussian to the temporal cross-correlation function and take the maximum as our best-fit phase offset. We then calculate the error on the Gaussian mean, considering the error in our correlations. The best-fit phase lags, $\Delta\phi$, and corresponding errors are listed in the third column of Table~\ref{phaseShifts}. These errors only reflect the photon-limited errors for each parameter, and may be an underestimate due to possible systematic noise terms, but do provide a rough assessment of the confidence level in each lag measurement. The GP model is more adept at modeling the error using the data itself. Given that our analytical model lags agree within 1-$\sigma$ of the GP-derived shifts, we adopt the GP-computed shifts for the remainder of our detrending analysis.

\begin{table}[!htb]
    \caption{Gaussian process (GP) and interpolative model (IM) derived phase lags between the line-by-line integrated RV time series and each activity indicator. Lag estimates derived from the IM all fall within the error-bars on the GP lags, and show a clear maximal correlation at 4-7 days.} 
    \centering
    \begin{tabular}{c c c}
        \hline
        Metric & $\Delta\phi_{\mathrm{LBL, GP}}$ (days) & $\Delta\phi_{\mathrm{LBL, IM}}$ (days) \\
        \hline\hline
        CaIIHK Index & $6.66 \pm 0.56$ & $6.39 \pm 0.12$\\ 

        H$\alpha$ Index & $6.67 \pm 0.64$ & $6.31 \pm 0.13$\\

        CaIRT Index & $5.31 \pm 0.62$ & $6.14 \pm 0.12$\\
        
        CCF FWHM & $4.65 \pm 1.64$ & $6.00 \pm 0.19$ \\
        
        CCF Contrast & $4.65 \pm 1.64$ & $5.73 \pm 0.20$ \\
        
        CCF BIS & $6.39 \pm 0.79$ & $6.02 \pm 0.43$ \\
        
        Depth Metric & $5.28 \pm 1.73$ & $4.87 \pm 0.30$\\
        
        \hline
    \end{tabular}

    \label{phaseShifts}
\end{table}

\subsubsection{Detrending bulk RVs using measured lag}

\begin{figure*}[!htb]
\includegraphics[width=\textwidth]{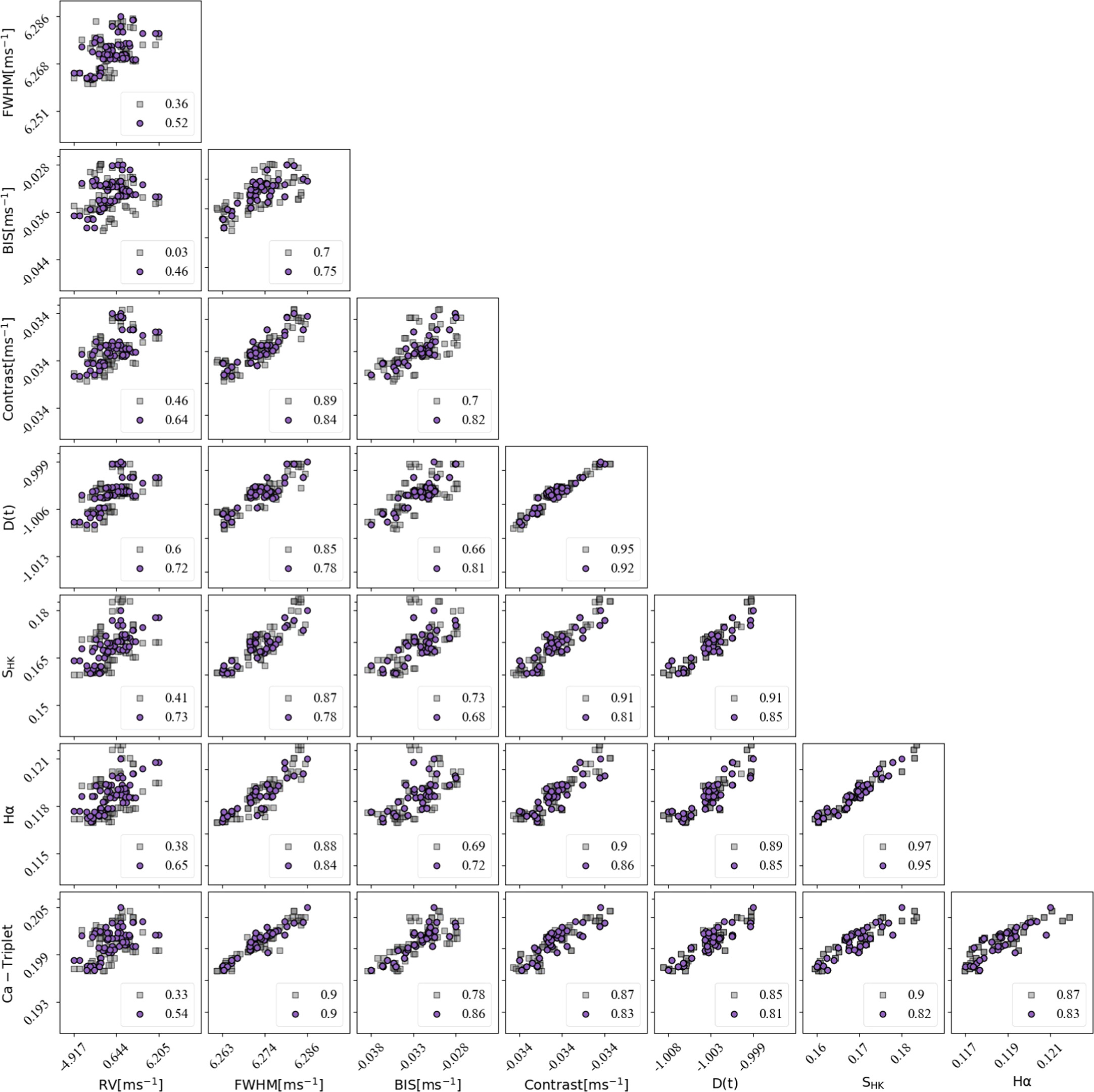}
\caption{Correlations between bulk RV and each activity metric described (described in Section~\ref{sec:bulkCorrs}). The two time series are correlated and plotted before (grey) and after (purple) being phase-shifted to maximize correlation with bulk RV. Correlation coefficients for both sets shows general improvement in correlation after shifting each indicator by its best fit GP modeled phase lag.}
 \label{fig:actIndsShifted}
\end{figure*}

To investigate the effect of phase shifting on the correlation between the two time series, we use the GP-derived lags to shift individual activity indicators to their maximal correlation phase. Once shifted, we linearly interpolate the DRP activity indicator measurements onto the same observation times as the RVs. We then recompute the correlation coefficients, excluding the first several days of data points, as the interpolation could be unreliable outside of the existing observation time window. The phase-shifted plots of all activity indicators are shown in Figure~\ref{fig:actIndsShifted}. The phase-shifted $S_\mathrm{HK}$ and D(t) indicators both show strong correlation with respective values of $R=0.73$ and $R=0.72$, while the CCF BIS shows the lowest correlations of $R=0.46$ (all improved from the un-shifted time series, as expected).

\begin{figure*}[!htb]
\centering
\includegraphics[width=0.85\textwidth]{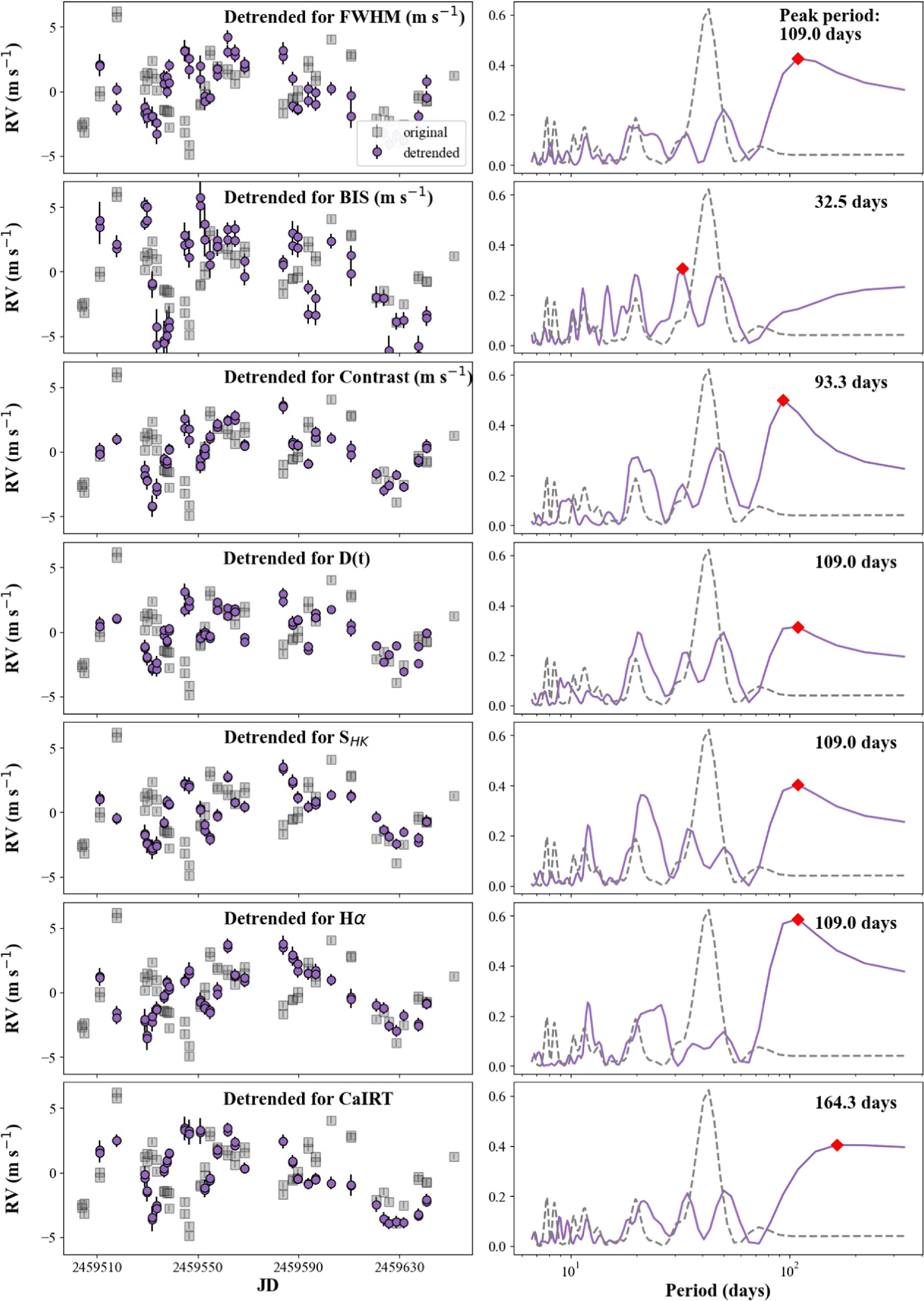}
\caption{\textit{Left}: Line-by-line bulk RV time series when detrended (purple circles) for each metric based on correlations shown in Figure \ref{fig:actIndsShifted}. The original bulk RVs are show in grey squares. \textit{Right}: Original RV periodogram (grey dashed line) plotted under detrended periodogram (purple solid line). The majority of detrended time series show substantial decrease in the relative power at or near the 42.6 day rotation period. For the line index and depth metric cases, power reduction ranges between 57.6$\%$ when detrending for the CaIR triplet and 82.99$\%$ when detrending for $S_\mathrm{HK}$. Note the first few days of observations are omitted in this analysis to avoid potential issues with extrapolation. The period with the highest remaining power is noted by the red diamond.}
\label{fig:actIndShiftDetrendPeriodograms}

\end{figure*}

Given these increased correlations, we investigate the periodicity in the residuals when detrending the RV time series for these phase shifted correlations. Figure \ref{fig:actIndShiftDetrendPeriodograms} shows the detrended RV time series and resulting periodograms. When detrending against the shifted activity indicators, we find a measurable reduction in the overall RV signal amplitude, as well as a strong depletion of the signal at the 42 day rotation period (Figure \ref{fig:actIndShiftDetrendPeriodograms}, right column). This strongly implies the observed velocity signal is activity-driven.

\subsection{Individual line RV analysis}\label{sec:lineRVAnalysis}

In addition to the bulk RV analysis, we investigate the potential activity signatures using individual line-by-line RVs. Similar to \cite{Cretignier_2020}, we explore the depth-sensitivity of the derived RV signals, and ultimately aim to understand if there is a strong depth dependence, indicative of an activity-induced signal driven by suppression of convective blueshift.

\begin{figure*}[!htb]
\centering
\includegraphics[width=0.9\textwidth]{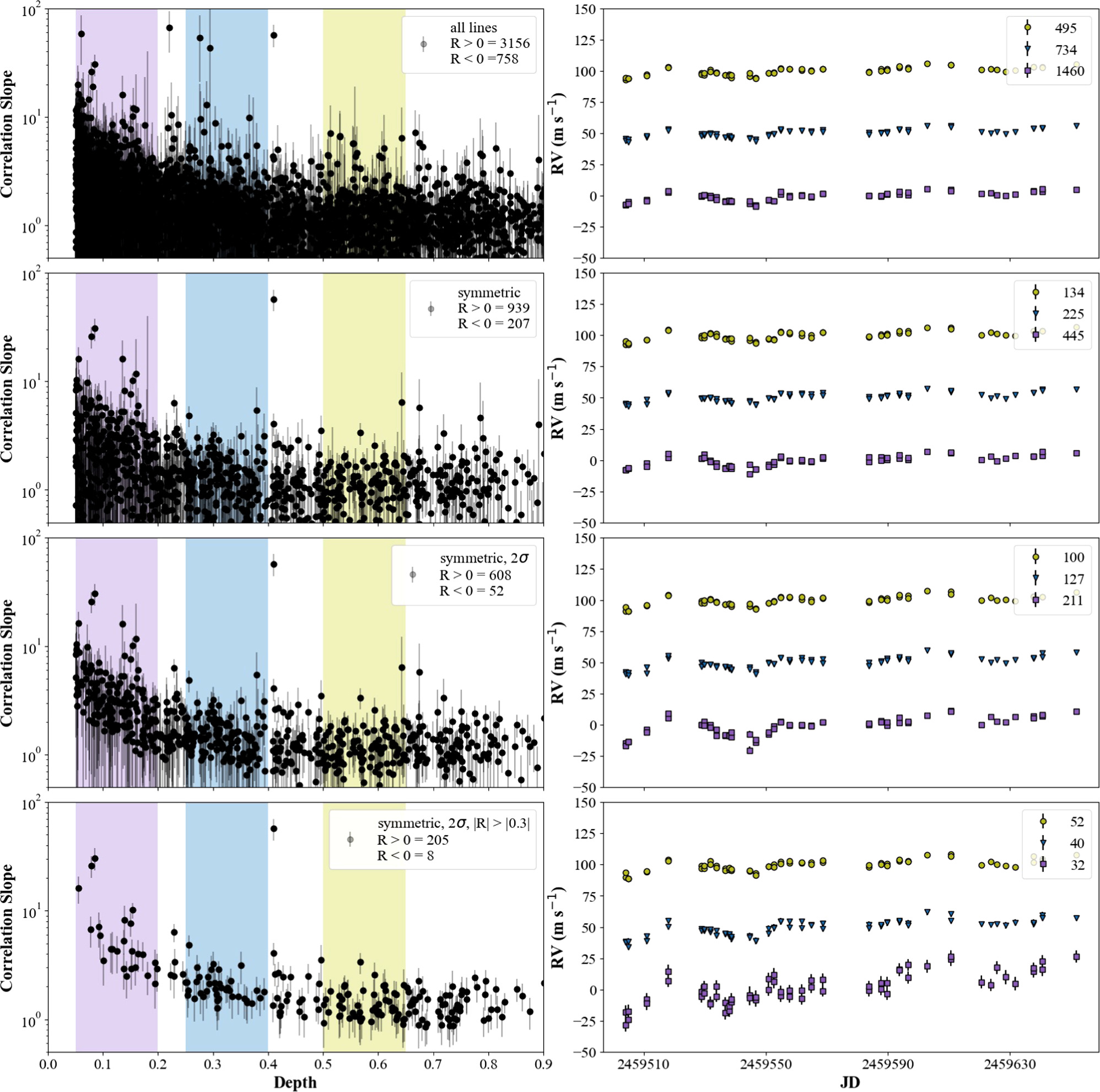}
\caption{\textit{Left:} Bulk-RV correlation slope (y-axis) versus line depth (x-axis) for different line selection populations (left to right): all lines with a depth greater than 0.05, symmetric lines, symmetric lines with $|2\sigma|$ confidence in the measurement of their correlation, and lines which additionally correlate with bulk-RV with a Pearson's R coefficient of $|R| \geq |0.3|$. Though the log plots show only the positively correlated lines, the integrated RV signals do include negatively correlated lines that meet the above thresholds (see legends). \textit{Right:} The corresponding depth-binned integrated RV time series for the populations described above, offset for clarity. All integrated RVs seem to share share a similar signal. The legends denote how many lines were used to compute the RVs in each depth bin. The depths bins are chosen to yield comparable overall RV uncertainty in each depth bin (1-3 {\ms} photon-limited error). The bins are 0.05 - 0.2, 0.25 - 0.4, and 0.5-0.65 in normalized depth.}
 \label{fig:eightPanelDepthRV}
\end{figure*}

\subsubsection{Line selection: bulk RV correlation}\label{sec:lineBulkCorrs}
Following the prescription of \cite{Cretignier_2020}, we select \lq{}high activity\rq{} lines based off of their correlation with the bulk RV signal. The underlying assumption of this approach is that the bulk RV signal is driven by activity effects. We correlate each line RV with our bulk line-by-line RV. 

To estimate error bars for our linear correlation coefficients, we remeasure the Pearson's linear correlation coefficient $R$ value 100 times, each time drawing a new value for the line RV and the bulk RV within their respective 1$\sigma$ error bars, computed by propagating photon-limited uncertainties. We then measure the width of the distribution of the resulting correlation coefficients values, which sets the 1-$\sigma$ error on $R$. To isolate lines with statistically significant signals, we consider only lines where the absolute value of the estimated correlation coefficient, $|R|$, exceeds twice the estimated measurement uncertainty i.e. ($|R|/{\sigma_R}$>2). For the high-bulk RV correlation population, we make a cut at a Pearson's R Coefficient of $|R| \geq |0.3|$. After applying these cuts, in addition to imposing the symmetry criteria described previously, we are left with 429 lines out of the original 6592.

Starting with the population of lines that shows strong correlation with the bulk RV signal (far right panel of Figure \ref{fig:eightPanelDepthRV}, we find a clear increase in the slope of the correlation between the individual line RV time series vs. bulk RV time series as a function of normalized line depth. This is broadly consistent with results shown in \cite{Cretignier_2020} for $\alpha$ Cen B. We then bin the RVs by line depth using three subsets of lines at increasing depth: 0.05 - 0.2 (shallowest), 0.25 - 0.4, and 0.5 - 0.65  (deepest). We find that the integrated RVs of each bin have a similar shape, but show significantly different amplitudes. This implies the signal is not planetary in nature, as described in \cite{Cretignier_2020}. If the signal were due to the proposed planet candidate signal, it would induce a signal of constant amplitude across all lines. Periodic suppression of convective blueshift, on the other hand, systematically affects some lines more strongly than others \citep{Cretignier_2020}.

\subsubsection{Exploring line selection bias}\label{sec:selectionBias}

Selecting lines by their correlation with the bulk RV alone may be subject to a systematic selection bias, which would be enhanced for shallow lines in particular. Given that shallow lines have a lower velocity information content, in order to have a stronger correlation with the bulk RV, that correlation must definitionally have a higher slope to achieve the same correlation coefficient. 

To verify that this potential bias does not significantly affect our selection,  we instead \lq{}blindly\rq{} select our lines based first on line symmetry (see previous section), and second on 2$\sigma$ certainty ($|R|/{\sigma_r}$>2) on the line-bulk RV correlation (as opposed to making a blanket cut using correlation with bulk RV). Both parameters are unrelated to time-series correlation with the bulk RV. Additionally, we expect that more symmetric lines will present RV time series with fewer spurious RV detections such as those induced by the movement of unresolved blends.

Since we are filtering based on the statistical certainty in the absolute correlation ($|R|/{\sigma_r}$>2), we do not reject anti-correlated lines. Using this selection process, we are left with 52 negatively correlated lines out of the total 660 lines that passed both criteria. We suspect that these are particularly subtle blends that made it through our symmetry cuts. Though the negative lines make up approximately 8$\%$ of the total line population, this is still a significant reduction compared to the unfiltered line list, where approximately 20$\%$ of lines are negatively correlated (758 / 3914). Were we to reject negatively correlated lines based on the assumption that they are due to subtle blends, we would only increase the amplitude of the depth-bin integrated RV time series.

\subsubsection{Depth Metric}
Since a single spectral line possesses only a small amount of information, \cite{Siegel2022} introduced the depth metric, $\mathcal{D}(t)$. For a given stellar spectrum, the depth metric is the weighted average line depth over a selected subpopulation of activity sensitive lines; the weighted average is calculated analogously to Eqn~\ref{eqn:rv_weighted_avg}. All lines satisfying $\mathcal{R}<5$th~percentile of HD26965's $\mathcal{R}$ distribution---where $\mathcal{R}$ is the Pearson correlation coefficient between a given line's depth time series and the bulk RV time series---were flagged as activity sensitive. For HD26965 the depth metric closely tracks the quasi-sinusoidal variations seen in the NEID RV time series, similar to the behavior observed in $\alpha$ Cen B \citep{Siegel2022}.

\subsubsection{FF' modeling}
\label{sec:ffprime}
\begin{figure*}[htb!]
\centering

\includegraphics[width=\textwidth]{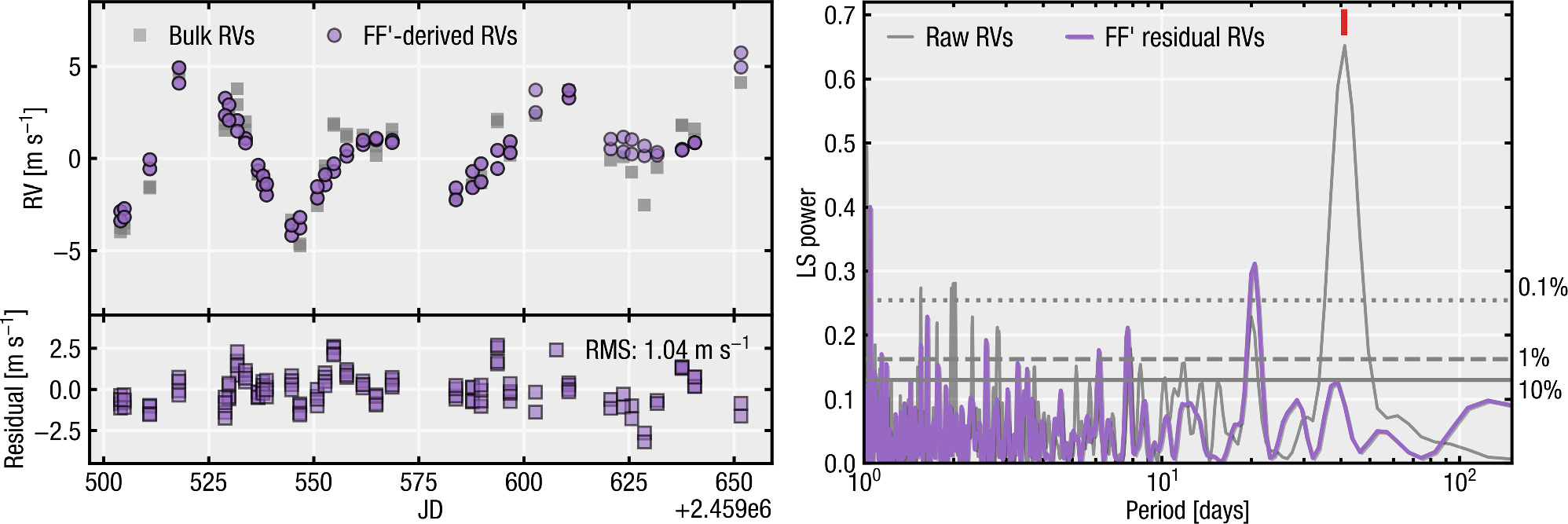}
\caption{\textit{Left}: Line-by-line bulk velocities with corresponding FF'-based velocities derived using the Depth Metric, $\mathcal{D}(t)$ \citep{Siegel2022}, to create corresponding theoretical activity signal (see Section~\ref{sec:ffprime}). Typical velocity error bars are smaller than the plotting symbols ($\sim$30-50 {\cms}). The RMS of the residuals is 1.04 {\ms}. \textit{Right}: Lomb-Scargle periodogram of the bulk RVs (gray) and FF' (purple) residuals, showing a significant reduction in the power at the $\sim$42 day candidate period (red dash). False alarm probability thresholds are shown in the horizontal lines for 0.1\% (dotted), 1\% (dashed), and 10\% (solid). We do note that a potential $P_\mathrm{rot} / 2$ alias may remain ($\sim$21 days) after subtraction of the best-fit activity signal, but it has a comparably low significance.}
\label{fig:ffprime_rvs}
\end{figure*}

Following \cite{Siegel2022}, we leveraged the depth metric time series  $\mathcal{D}(t)$ to detrend the RV time series of HD26965 using a modified FF$^{\prime}$ model. First described by \cite{Aigrain2012}, the FF$^{\prime}$ model uses photometric flux and an active region model to predict the RV signal due to stellar activity. \cite{Giguere2016} then introduced the HH$^{\prime}$ method, which uses the H$\alpha$ index as a photometric proxy. Using a modified FF$^{\prime}$ framework, we treated the relative amplitudes of the spot rotation and convective blueshift effects as a free parameter and allowed for a linear relationship between $\mathcal{A}(t)$ and flux \citep{Siegel2022}:
\begin{align}
    \label{eqn:ffprime}
    \text{FF}^{\prime}(t) = &-\alpha(\mathcal{A}(t) +\beta)\dot{\mathcal{A}}(t)/f  \nonumber \\
    &+ \gamma (\mathcal{A}(t)+\beta)^2 /f \nonumber \\
    &+ \mathcal{C}_2 t + \mathcal{C}_1,
\end{align}
where $\mathcal{A}(t)$ is a normalized activity index time series (in this case,  $\mathcal{D}(t)$), $f$ is the relative flux change for a feature at the disk center (included to conveniently scale $\alpha$ and $\gamma$), $\beta$ is the zero-point of the assumed linear relationship between $\mathcal{A}(t)$ and photometric flux, and $C_1$ and $C_2$ are an arbitrary zero-point and linear drift, respectively; $f$ is defined analogously to \cite{Aigrain2012}.
To infer $\dot{\mathcal{A}}(t)$, $\mathcal{A}(t)$ is smoothed using a smoothing parameter $\sigma_t$ \citep{Siegel2022}. For the $j$th observation, the smoothed $\mathcal{A}_j$ is the weighted average of the entire activity metric time series, where the weights are assigned via a Gaussian centered at $t_j$ (the time of the $j$th observation) with a standard deviation of $\sigma_t$. The smoothed activity metric is fit with a cubic spline and $\dot{\mathcal{A}}(t)$ is determined analytically.
The FF$^{\prime}$ RVs are given by Eqn.~\ref{eqn:ffprime}, which has six free parameters: $\alpha$, $\beta$, $\gamma$, $\mathcal{C}_1$, $\mathcal{C}_2$, and $\sigma_t$. The model was optimized via affine-invariant MCMC sampling using \texttt{emcee} \citep{ForemanMackey2013}. Uniform priors were adopted for the activity model parameters. We employed 90 walkers for 5,000 iterations each, where the first 1,000 iterations were rejected as burn-in. The resulting FF'-derived RVs are shown in Figure~\ref{fig:ffprime_rvs}.

These RVs show that the bulk of the observed RV signal can be modeled with a relatively simple activity-driven framework, and the residual signal shows significantly less power at the 42 day period (see Figure~\ref{fig:ffprime_rvs}, right).  {It is worth noting that any keplerian signals that are 1) in-phase with the stellar activity signal, and 2) have periods that are identical (or very near) the stellar rotation period will always be partially subtracted in this framework due to model degeneracy (see Appendix) A potential solution to this would be to add additional information into the model, such as including differential wavelength information (e.g. simultaneously fitting the blue and red wavelengths, and positing the Keplerian signal should be independent of color while the activity signal will be color dependent). As the claimed planet is significantly out of phase with the observed activity cycle (Figure~\ref{fig:MaModel}), we are confident that our model is not significantly reducing this specific signal.}

\subsection{Implications for planet detectability} \label{sec:planet_detetability}

Previous studies of this target have found linear detrending against activity signals  ineffective \citep{Ma2018, Diaz_2018}. The inclusion of a phase shift, as outlined above, may explain why these studies have not identified the activity-to-RV correlation at the 42 day period. This could also be due to a difference in the overall activity level of {\target}, which may explain why the correlation amplitudes are stronger in our NEID dataset than in previous HARPS observations \citep{Ma2018}.

We also note that previous investigations into {\target} did not explore the variation in the planet candidate signal in different line populations parsed by symmetry, depth, or activity-correlation. Variable velocity amplitudes at or near the rotation period in these various line samples strongly suggest that the potential Keplerian near the rotation period is in fact driven by activity.

\subsubsection{Best-fit Keplerians}

We use the \texttt{radvel} package \citep{Fulton2018} to find the best fit Keplerian orbits and corresponding errors, fitting for a single planet signal. We compute Keplerian fits on the residuals of the detrended RV time series of Figure \ref{fig:actIndShiftDetrendPeriodograms}. As shown in Table \ref{tab: scale}, we find that although the best fit periods agree to within 1- to 2-sigma, the RV semi-amplitude varies between 1.0 and 2.35 ms$^{-1}$. The best fit amplitude of around 1 m s$^{-1}$ for the CaHK, H$\alpha$, Ca IR triplet, and Depth metric indicators reveals that the RV residuals post-detrending would be sensitive to the 1.8 m s$^{-1}$ signal claimed in \citet{Ma2018}, but this signal was not recovered in the activity-detrended NEID data.

\begin{figure*}[!htb]
\centering
\includegraphics[width=0.8\textwidth]{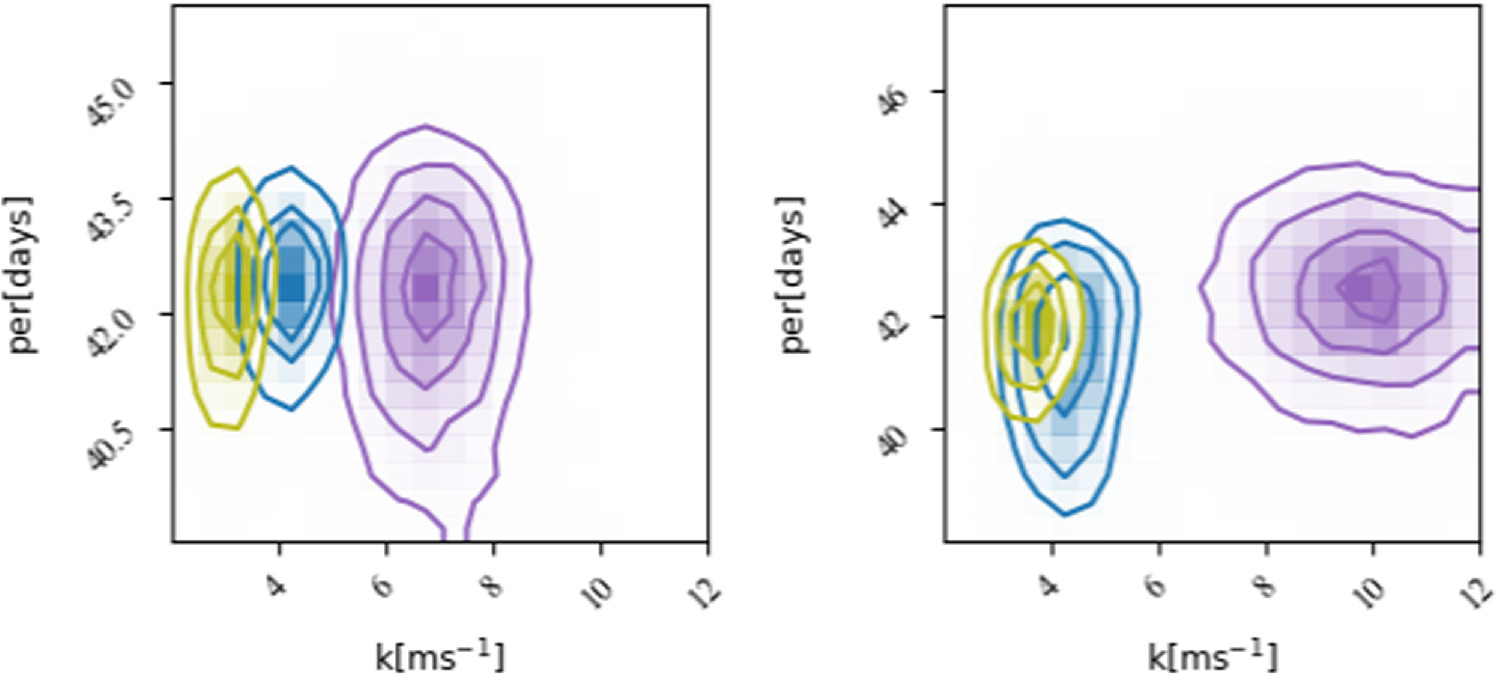}

\caption{Posteriors from MCMC best-fit Keplerian periods, per, and semi-amplitudes, $K$, for the symmetric, 2-$\sigma$ lines (left) and the symmetric, 2-$\sigma$, $R > 0.3$ lines (right). Fits for each depth bin highlighted in Figure~\ref{fig:eightPanelDepthRV} are shown in yellow for deep lines, blue for mid-depth lines, and purple for shallow lines. The contour lines mark the 68\%, 95\%. and 99.7\% confidence limits. The recovered planet semi-amplitude ($K$) varies significantly as a function of line depth in both cases.}
    \label{fig:Posteriors}
\end{figure*}

Similarly, we fit separate Keplerian signals to the shallow and deep line RV time series using the different line populations shown in Figure \ref{fig:eightPanelDepthRV}. We find that the best fit RV semi-amplitude varies significantly as a function of line depth, which is inconsistent with a Keplerian signal but expected if the observed RVs are dominated by periodic suppression of convective blueshift \citep{Cretignier_2020}. Figure~\ref{fig:Posteriors} shows the Monte Carlo posteriors of the best-fit Keplerian for our three line depth bins in both the symmetry selected and activity correlation selected line populations. In both populations, the 1-$\sigma$ best-fit amplitudes of the shallow line bins are over two times higher than the deep line bins. Given that we find neither any consistent best-fit Keplerian orbit nor the specific signal discovered by \citet{Ma2018}, we believe that the large amplitude differences between our best-fit signals further suggest that the variability in the {\target} RVs is driven by activity.

\begin{figure*}[htb]
\centering
\includegraphics[width=\textwidth]{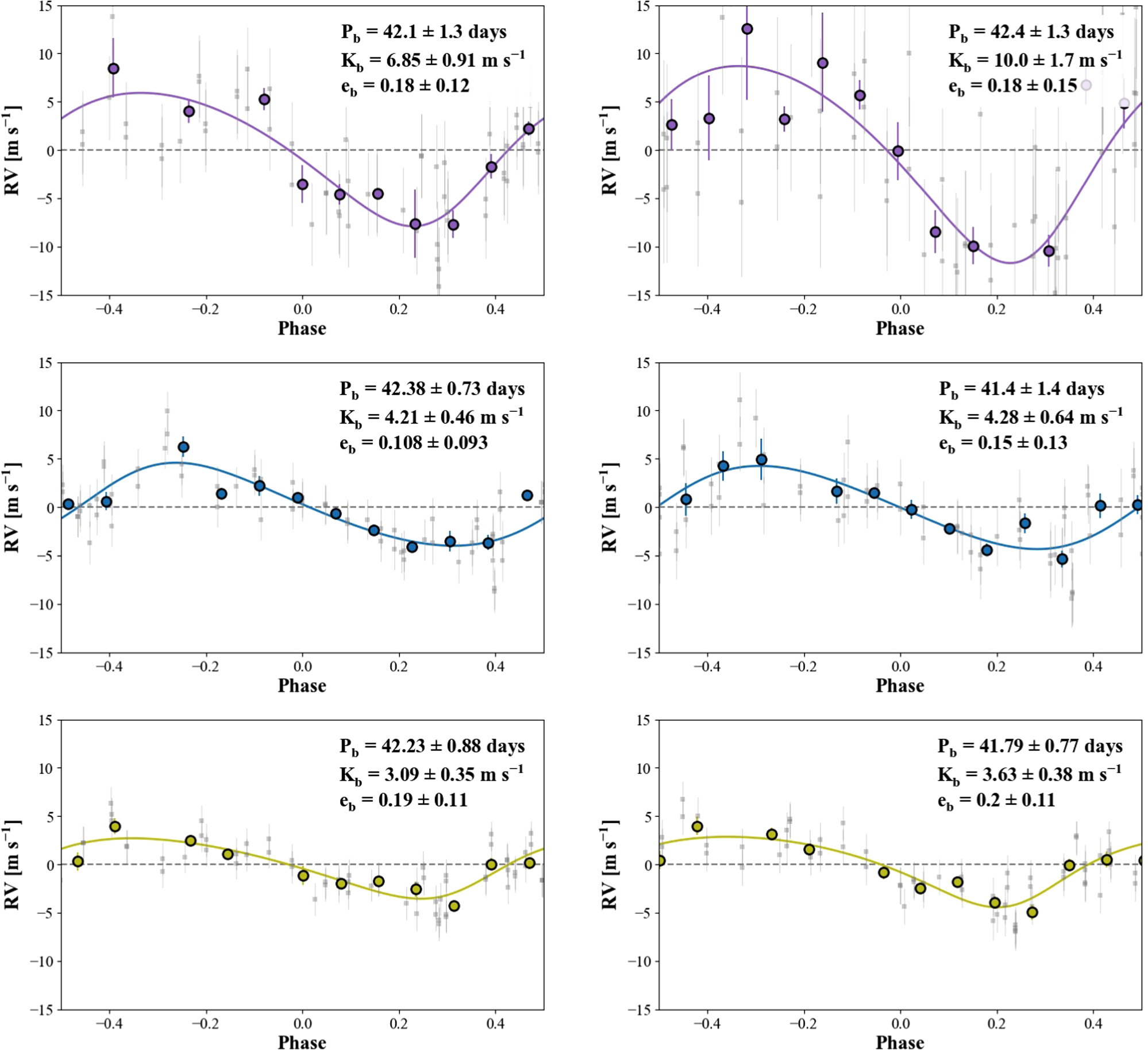}
\caption{Phase-folded velocity plots of the best-fit Keplerians corresponding to the MCMC fits of Figure~\ref{fig:Posteriors} for the symmetric, 2-$\sigma$ lines (left) and the symmetric, 2-$\sigma$, $R > 0.3$ lines (right). Median phase-binned points for each depth bin are in yellow for deep lines, blue for mid-depth lines, and purple for shallow lines. Unbinned, phase-folded RVs for each depth bin are shown in grey. Colored curves show the best-fit $\tt \texttt{radvel}$ models.}
\label{fig:phasedPlots}
\end{figure*}

\begin{table}
    \caption{\textit{Top}: Best-fit planetary signal for metric-detrended RVs. While a signal  between 40-43 days is consistently recovered, the wide range in fitted amplitudes implies that the underlying RVs are driven by activity. \textit{Bottom}: Best-fit planetary signal for line-selected integrated RV signals from Figure~\ref{fig:eightPanelDepthRV}. Though we recover the near 42-day signal in each line population, the amplitude of the signal increases by nearly a factor of two from the deepest to shallowest line populations.}  
    \begin{centering}
    \begin{tabular}{c c c}
        \hline\hline
        Metric & P (days) & K (m s$^{-1}$) \\
        \hline
        CCF FWHM & $42.59 \pm 0.55$ & $2.04 \pm 0.20$\\
        
        CCF BIS & $40.78 \pm 0.71$ & $2.35 \pm 0.31$\\
        
        $S_\mathrm{HK}$ & $41.72 \pm 0.99$ & $1.0 \pm 0.09$\\
        
        H$\alpha$ & $42.26 \pm 0.92$ & $1.0 \pm 0.08$\\
        
        CaIRT & $42.26 \pm 0.87$ & $1.23 \pm 0.20$\\
        
        Depth Metric & $41.87 \pm 0.93$ & $1.0 \pm 0.09$\\

        \hline
        
        Shallow lines & $42.1 \pm 1.3$ & $6.85\pm 0.91$\\

        Mid lines & $42.38 \pm 0.73$ & $4.21 \pm 0.46$\\
        
        Deep lines & $42.23 \pm 0.88$ & $3.09 \pm 0.35$\\
        \hline
    \end{tabular}
    \end{centering}

    \label{tab: scale}
\end{table}

\section{Conclusion} 
\label{sec:conclusion}
Employing a combination of activity indicators and line-by-line techniques in NEID data has shown that the likely origin of the periodic signal seen in {\target} is stellar activity. We have a relatively short baseline of observations, but the line-by-line signals clearly show a statistically significant variation in the periodic signal amplitude. We conclude that this variation is likely indicative of activity and not Keplerian motion. 

We show that classical activity indicators, such as CCF FWHM and various line indices ($S_\mathrm{HK}$, H-$\alpha$, etc.) all show similar rotational modulation as {\target}'s integrated RVs. We investigated the potential phase shift between these indicators and the RVs, due to the delay between the line morphology, equivalent width, and RV effects of stellar active regions. We found consistent phase shifts for the line indicators of between 5 and 8 days (42 - 68 degrees in orbital phase, relative to the rotation period) when fitting the phase using both a GP and a basic interpolated model. We verified that phase shifting the activity metrics to the same phase as the RVs not only boosted their correlations but largely depleted the periodogram power at the 42.4 day period in the RVs when linearly detrending. This depletion shows that phase offsets may be key in identifying and characterizing the relationship between radial velocities and underlying activity.

Using our line-by-line RV pipeline on {\target}, we were able to recover a similar result to  {\citet{Dumusque-2018}} on $\alpha$ Cen B, which posits that the RVs of shallow lines are subject to the convective blueshift of active regions and thus show larger RV variations. We selected a group of 552 symmetric lines with 2-$\sigma$ correlations with the bulk RV and demonstrated that the best fit Keplerian of the combined RV signals of the shallow lines not only favor a higher amplitude signal than that of the deep lines, but also show a greater amplitude than the best-fit Keplerian orbit from \citet{Ma2018}. 

We theorize that the efficacy of detrending radial velocities for phase-shifted activity indicators points towards a decaying star spot or plage with a phase lag between its RV strength and magnetic strength \citep{CollierCameron-2019}. Though we acknowledge that detrending exercises may overstate the influence of one parameter on the other, we argue that the existence of a semi-consistent phase-shift between multiple activity indicators with identical periodicity to the integrated radial velocities strongly points towards a direct relationship between activity and the observed RV signal.

We additionally find a line-depth dependent velocity amplitude at the same period, which is in line with the expected effect of the inhibition of convective blueshift in active regions. This explanation is broadly consistent with the decaying star spot or plage scenario outlined above.
 {While each of these methods taken individually may not rule out a potential planetary signal at the same phase and period as the activity signal, collectively our analyses show that an activity hypothesis is favored over the specific planet claimed in \citep{Ma2018}.}
In the future, we aim to explore more exotic, physically-motivated methods for grouping spectral lines when computing integrated stellar RVs. These methods include exploring integrated RVs as a function of formation temperature \citep{AlMoulla2022}, normalized depth \citep{Siegel2022}, and various other fundamental parameters such as excitation potential and Land\'e g-factor \citep{Wise-2018}. Computing other empirical line parameters, such as equivalent width, skewness, and bisector may also prove to be valuable for exploring activity correlations for individual features.

We also hope to apply these analyses to Solar data during periods of heightened activity, with the goal of discerning which lines are better tracers of specific activity phenomena. Directly comparing individual spectral line parameters such as depth, asymmetry, and velocity as well as classical activity indicators to pre-computed solar parameters from the Solar Dynamics Observatory (SDO) may shed light on this \citep{Ervin2022}. Additionally, NEID Solar data and SDO images could also be combined to further test both the active region phase-lag and convective blueshift hypothesis proposed here.Finally, extending our approach to other spectral types, or even similar spectral types at a range of ages, could be an interesting test of the universality of our methods across the main sequence.

\section*{Acknowledgements}
Data presented were obtained by the NEID spectrograph built by Penn State University and operated at the WIYN Observatory by NOIRLab, under the NN-EXPLORE partnership of the National Aeronautics and Space Administration and the National Science Foundation. Based in part on observations at the Kitt Peak National Observatory, managed by the Association of Universities for Research in Astronomy (AURA) under a cooperative agreement with the National Science Foundation. WIYN is a joint facility of the University of Wisconsin–Madison, Indiana University, NSF’s NOIRLab, the Pennsylvania State University, Purdue University, University of California, Irvine, and the University of Missouri. The authors are honored to be permitted to conduct astronomical research on Iolkam Du’ag (Kitt Peak), a mountain with particular significance to the Tohono O’odham. Data presented herein were obtained from telescope time allocated to NN-EXPLORE through the scientific partnership of the National Aeronautics and Space Administration, the National Science Foundation, and the National Optical Astronomy Observatory. The research was carried out at the Jet Propulsion Laboratory, California Institute of Technology, under a contract with the National Aeronautics and Space Administration (80NM0018D0004) and funded through the President’s and Director’s Research \& Development Fund and Research and Technology Development Programs. 
E.B.F.'s contributions were partially supported by Heising-Simons Foundation Grant \#2019-1177, NSF AST award \#2204701 and NASA EPRV award \#80NSSC21K1035.
The Center for Exoplanets and Habitable Worlds is supported by the Pennsylvania State University and the Eberly College of Science.
GS acknowledges support provided by NASA through the NASA Hubble Fellowship grant HST-HF2-51519.001-A awarded by the Space Telescope Science Institute, which is operated by the Association of Universities for Research in Astronomy, Inc., for NASA, under contract NAS5-26555.

\software{
\texttt{astropy} \citep{AstropyCollaboration2018_astropy},
\texttt{barycorrpy} \citep{Kanodia2018_barycorrpy}, 
\texttt{matplotlib} \citep{Hunter2007_matplotlib},
\texttt{numpy} \citep{vanderWalt2011_numpy},
\texttt{pandas} \citep{McKinney2010_pandas},
\texttt{scipy} \citep{Virtanen2020_scipy},
}

\appendix 

\label{sec:appendix}\section{FF' model Injection recovery test}  {To test the robustness of our modified FF' model (Equation~\ref{eqn:ffprime}) specifically for signals around the stellar rotation period,  we conduct an injection test of fitting the model to sample Keplerian sinusoids. We first inject known orbits into the raw LBL RVs at a variety of different phases relative to the observed activity signal. We then apply the FF' model and fit the residual RVs to attempt to recover the injected planet signal. We test two different orbital periods near the stellar rotation period (32 days and 42 days), each at five different orbital phases relative to the larger activity signal. For each choice of period and phase, we fit the RVs with the FF' model using H-alpha as the flux proxy (A(t)) to derive the best-fit activity-induced signal.}

\begin{figure*}[htb!]
\centering

\includegraphics[width=0.9\textwidth]{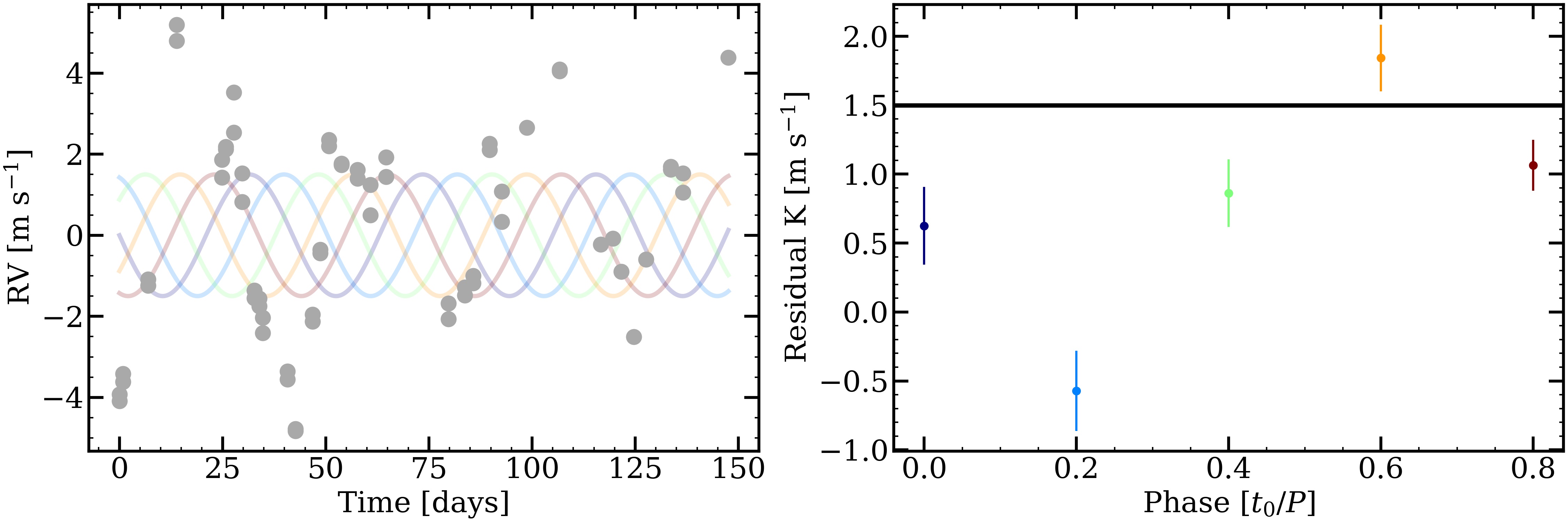}
\includegraphics[width=0.9\textwidth]{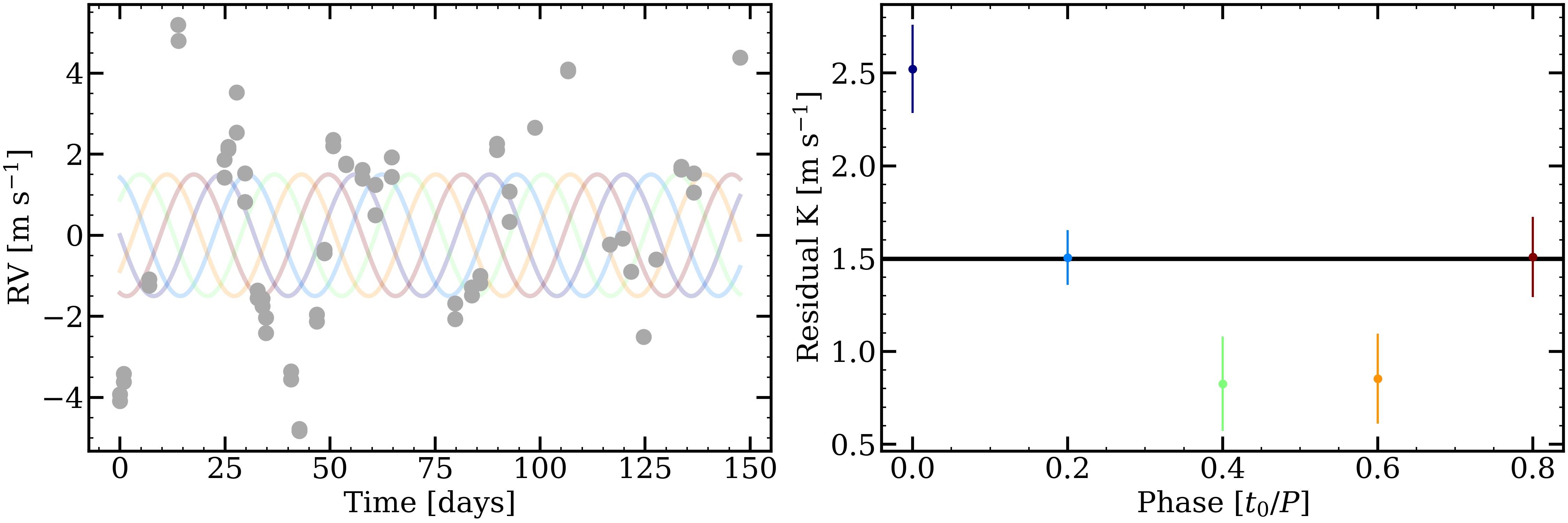}
\caption{ {Overview of FF' injection recovery experiment, applied to the measured bulk RVs of {\target}. A 1.5 {\ms} signal was injected at two different periods: 32 days (top) and 42 days (bottom). Each period sampled five different relative phases ($t_{0}/P$) between the keplerian signal and the stellar rotation / activity signal. Our FF' model was then applied to the combined RV data, and a single keplerian was fit the residuals. The recovered semiamplitudes ($K$) are shown in the right column. In the case where the phase of the planet is at or near the activity, the recovery of the planet signal is significantly degraded when relying on the FF' model alone, while injected signals that are out-of-phase with the activity are generally recovered more reliably.}}
\label{fig:injrecovery_test}
\end{figure*}

 {The right panels in Figure~\ref{fig:injrecovery_test} show the resulting RV semi-amplitudes fit to the residuals after the activity subtraction. When the planet's signal is out of phase with the activity signal, the recovered K is close to the injected K. When the planet's signal is in phase with the activity signal, the planet signal is obscured. This is unfortunately a guarantee when the activity signal is near sinusoidal and does not change much over the observing baseline (which is the case here, since the activity amplitude and period are nearly constant over the 150 day observation window); in cases like this, the planet and activity signals are highly degenerate when their phases are closely aligned. As such, for this study we use the FF' modeling as only one piece of evidence casting doubt on the the \citep{Ma2018} candidate which, when combined with other activity detrending approaches, provide strong evidence that the planet candidate is indeed likely attributable to activity alone.}

\bibliography{main}{}
\bibliographystyle{aasjournal}

\startlongtable

\tabletypesize{\scriptsize}

\begin{deluxetable}{lrrrrrrrrrl}

\tablewidth{0pc}
\tablecaption{
    Summary table of bulk NEID measurements of {\target} used in this study, including bulk velocities, CCF activity indicators, \lq{}classical\rq{} line activity indicators, and the computed Depth Metric ($\mathcal{D}(t)$)
    \label{tab:rvs}
}
\tablehead{
    \colhead{JD} &
    \colhead{RV$_\mathrm{NEID}$} &
    \colhead{RV$_\mathrm{LBL}$} &
    \colhead{$\Delta$CCF$_{\mathrm{FWHM}}$} &
    \colhead{$\Delta$CCF$_{\mathrm{BIS}}$} &
    \colhead{$\Delta$CCF$_{\mathrm{Contrast}}$} &
    \colhead{($\mathcal{D}(t)$ + 1))} &
    \colhead{$S_\mathrm{HK}$} &
    \colhead{H$\alpha$} &
    \colhead{CaIRT} \\
    \colhead{- 2459500} &
    \colhead{({\ms})} &
    \colhead{({\ms})} &
    \colhead{({\ms})} &
    \colhead{({\ms})} &
    \colhead{$\times10^{4}$} &
    \colhead{$\times10^{3}$} &
    \colhead{${\pm}$0.00055} &
    \colhead{${\pm}$0.00013} &
    \colhead{${\pm}$5.8e-5} &
}
\startdata
  4.019 &       -4.25 $\pm$ 0.23 &      -2.51 $\pm$ 0.27 &        -3.89 $\pm$ 0.04 &      -31.62 $\pm$ 0.08 & 342.694 $\pm$ 0.2 & -5.85 $\pm$ 0.007726 & 0.16099244 & 0.11706787 & 0.19839492 \\
   4.021 &       -4.25 $\pm$ 0.23 &       -2.7 $\pm$ 0.26 &        -8.27 $\pm$ 0.04 &      -33.01 $\pm$ 0.08 & 342.639 $\pm$ 0.2 & -6.07 $\pm$ 0.007738 & 0.16110582 & 0.11696177 & 0.19820225 \\
     4.9 &       -3.52 $\pm$ 0.22 &      -2.37 $\pm$ 0.26 &        -6.09 $\pm$ 0.04 &      -32.91 $\pm$ 0.08 & 342.618 $\pm$ 0.2 & -6.18 $\pm$ 0.007624 & 0.16061176 & 0.11670715 & 0.19721277 \\
   4.902 &       -3.81 $\pm$ 0.22 &      -3.18 $\pm$ 0.26 &        -6.44 $\pm$ 0.04 &      -32.75 $\pm$ 0.08 & 342.671 $\pm$ 0.2 & -6.18 $\pm$ 0.007624 & 0.16061021 & 0.11670716 & 0.19721174 \\
  10.971 &        -1.0 $\pm$ 0.23 &       0.04 $\pm$ 0.26 &        -7.42 $\pm$ 0.04 &      -34.52 $\pm$ 0.08 & 342.712 $\pm$ 0.2 & -6.08 $\pm$ 0.007675 & 0.16066467 & 0.11675316 & 0.19787301 \\
  10.974 &       -1.17 $\pm$ 0.23 &      -0.36 $\pm$ 0.26 &        -4.36 $\pm$ 0.04 &      -34.89 $\pm$ 0.08 &  342.64 $\pm$ 0.2 & -6.08 $\pm$ 0.007676 & 0.16066473 & 0.11675327 & 0.19787331 \\
   17.89 &        5.51 $\pm$ 0.23 &        6.2 $\pm$ 0.24 &         5.34 $\pm$ 0.04 &      -31.25 $\pm$ 0.08 & 342.067 $\pm$ 0.2 & -1.89 $\pm$ 0.007926 & 0.17008271 & 0.11916122 & 0.19981397 \\
  17.894 &        5.03 $\pm$ 0.23 &       5.78 $\pm$ 0.24 &         5.04 $\pm$ 0.04 &      -31.43 $\pm$ 0.08 & 342.064 $\pm$ 0.2 & -1.89 $\pm$ 0.007926 & 0.17008759 & 0.11916245 & 0.19981491 \\
  28.905 &        1.71 $\pm$ 0.23 &       1.21 $\pm$ 0.23 &         9.93 $\pm$ 0.04 &      -29.68 $\pm$ 0.08 & 341.641 $\pm$ 0.2 &  0.52 $\pm$ 0.007939 & 0.18265396 &  0.1212022 & 0.20347861 \\
  28.908 &        1.29 $\pm$ 0.23 &       0.14 $\pm$ 0.23 &        10.03 $\pm$ 0.04 &      -30.73 $\pm$ 0.08 & 341.554 $\pm$ 0.2 &  0.52 $\pm$ 0.007938 & 0.18265717 & 0.12120267 & 0.20347954 \\
  29.788 &        1.75 $\pm$ 0.23 &       0.87 $\pm$ 0.22 &         9.77 $\pm$ 0.04 &      -30.87 $\pm$ 0.08 & 341.555 $\pm$ 0.2 &  0.95 $\pm$ 0.007929 & 0.18340881 & 0.12151772 & 0.20393195 \\
  29.791 &        2.03 $\pm$ 0.23 &       1.48 $\pm$ 0.22 &        11.39 $\pm$ 0.04 &      -30.12 $\pm$ 0.08 & 341.548 $\pm$ 0.2 &  0.95 $\pm$ 0.007929 &  0.1834109 & 0.12151923 & 0.20393434 \\
  31.776 &        4.11 $\pm$ 0.23 &       2.37 $\pm$ 0.22 &         9.09 $\pm$ 0.04 &      -27.25 $\pm$ 0.08 & 341.349 $\pm$ 0.2 &  0.96 $\pm$ 0.007827 & 0.18293403 & 0.12069714 & 0.20476883 \\
  31.778 &        3.15 $\pm$ 0.23 &        1.4 $\pm$ 0.21 &         9.69 $\pm$ 0.04 &      -27.47 $\pm$ 0.08 & 341.359 $\pm$ 0.2 &  0.96 $\pm$ 0.007827 & 0.18293332 & 0.12069622 & 0.20476973 \\
   33.75 &        2.72 $\pm$ 0.23 &       0.99 $\pm$ 0.22 &         9.66 $\pm$ 0.04 &      -24.73 $\pm$ 0.08 & 341.563 $\pm$ 0.2 &  0.95 $\pm$ 0.007805 &  0.1798657 & 0.11984547 & 0.20355968 \\
  33.752 &         2.0 $\pm$ 0.23 &        0.1 $\pm$ 0.22 &         7.65 $\pm$ 0.04 &      -25.74 $\pm$ 0.08 & 341.618 $\pm$ 0.2 &  0.95 $\pm$ 0.007805 & 0.17986254 & 0.11984471 & 0.20355833 \\
  36.732 &       -0.32 $\pm$ 0.23 &      -1.46 $\pm$ 0.24 &         5.23 $\pm$ 0.04 &      -25.41 $\pm$ 0.08 & 342.012 $\pm$ 0.2 &  -2.0 $\pm$ 0.007819 & 0.17499944 & 0.11938667 & 0.20213142 \\
  36.735 &       -0.25 $\pm$ 0.23 &      -1.38 $\pm$ 0.24 &         3.73 $\pm$ 0.04 &      -25.26 $\pm$ 0.08 & 342.051 $\pm$ 0.2 &  -2.0 $\pm$ 0.007819 & 0.17499502 & 0.11938626 & 0.20213011 \\
  37.908 &       -0.24 $\pm$ 0.23 &      -1.57 $\pm$ 0.25 &         1.19 $\pm$ 0.04 &      -25.54 $\pm$ 0.08 & 342.196 $\pm$ 0.2 & -2.76 $\pm$ 0.007949 & 0.17264905 & 0.11906597 & 0.20125261 \\
  37.913 &       -0.61 $\pm$ 0.23 &      -1.43 $\pm$ 0.25 &         4.23 $\pm$ 0.04 &      -25.21 $\pm$ 0.08 & 342.239 $\pm$ 0.2 & -2.76 $\pm$ 0.007949 & 0.17263926 & 0.11906484 & 0.20124928 \\
  38.736 &       -1.59 $\pm$ 0.23 &      -2.77 $\pm$ 0.25 &        -0.92 $\pm$ 0.04 &      -26.47 $\pm$ 0.08 & 342.468 $\pm$ 0.2 &  -4.0 $\pm$ 0.007741 & 0.17109959 & 0.11893218 & 0.20068881 \\
  38.738 &        -1.1 $\pm$ 0.23 &      -1.56 $\pm$ 0.25 &        -1.36 $\pm$ 0.04 &       -26.2 $\pm$ 0.08 & 342.455 $\pm$ 0.2 &  -4.0 $\pm$ 0.007741 & 0.17109698 & 0.11893173 & 0.20068773 \\
  44.741 &        -4.01 $\pm$ 0.3 &      -3.19 $\pm$ 0.31 &        -9.22 $\pm$ 0.04 &      -33.06 $\pm$ 0.08 & 343.139 $\pm$ 0.2 & -8.59 $\pm$ 0.012201 & 0.16047432 & 0.11715399 & 0.19761635 \\
  44.749 &       -4.18 $\pm$ 0.27 &      -2.25 $\pm$ 0.29 &       -11.34 $\pm$ 0.04 &      -32.43 $\pm$ 0.08 & 343.016 $\pm$ 0.2 &  -8.6 $\pm$ 0.012206 & 0.16046163 & 0.11715198 & 0.19761265 \\
   46.69 &       -6.64 $\pm$ 0.23 &      -4.13 $\pm$ 0.27 &       -10.09 $\pm$ 0.04 &      -32.85 $\pm$ 0.08 & 343.095 $\pm$ 0.2 & -7.97 $\pm$ 0.007816 &  0.1598822 & 0.11693049 & 0.19741744 \\
  46.693 &       -6.45 $\pm$ 0.23 &      -4.92 $\pm$ 0.27 &        -9.65 $\pm$ 0.04 &      -32.01 $\pm$ 0.08 & 342.948 $\pm$ 0.2 & -7.98 $\pm$ 0.007812 & 0.15988115 & 0.11693014 & 0.19741696 \\
  50.848 &       -5.21 $\pm$ 0.23 &      -0.94 $\pm$ 0.26 &        -8.12 $\pm$ 0.04 &      -35.55 $\pm$ 0.08 & 343.033 $\pm$ 0.2 & -7.58 $\pm$ 0.007701 & 0.16152854 & 0.11731586 & 0.19722633 \\
   50.85 &       -5.06 $\pm$ 0.23 &      -1.06 $\pm$ 0.26 &        -8.41 $\pm$ 0.04 &      -36.13 $\pm$ 0.08 & 342.929 $\pm$ 0.2 & -7.58 $\pm$ 0.007701 & 0.16152914 &  0.1173161 & 0.19722616 \\
  52.715 &       -3.19 $\pm$ 0.23 &       0.41 $\pm$ 0.26 &        -7.42 $\pm$ 0.04 &      -34.82 $\pm$ 0.08 & 342.771 $\pm$ 0.2 & -6.14 $\pm$ 0.007804 & 0.16456582 & 0.11722842 & 0.19878021 \\
  52.718 &        -3.6 $\pm$ 0.23 &      -0.19 $\pm$ 0.26 &        -6.91 $\pm$ 0.04 &       -33.9 $\pm$ 0.08 & 342.856 $\pm$ 0.2 & -6.14 $\pm$ 0.007804 & 0.16457067 &  0.1172281 & 0.19878315 \\
  54.778 &        0.33 $\pm$ 0.23 &       2.82 $\pm$ 0.25 &        -2.72 $\pm$ 0.04 &      -32.69 $\pm$ 0.08 & 342.679 $\pm$ 0.2 & -3.92 $\pm$ 0.007768 & 0.16538283 & 0.11757045 & 0.19957359 \\
   54.78 &         0.1 $\pm$ 0.23 &       3.17 $\pm$ 0.25 &         -4.7 $\pm$ 0.04 &      -33.25 $\pm$ 0.08 & 342.705 $\pm$ 0.2 & -3.91 $\pm$ 0.007768 & 0.16538494 & 0.11757072 & 0.19957464 \\
  57.836 &         0.3 $\pm$ 0.23 &       1.97 $\pm$ 0.24 &        -0.09 $\pm$ 0.04 &      -32.22 $\pm$ 0.08 & 342.378 $\pm$ 0.2 &  -3.4 $\pm$ 0.007751 & 0.16773768 & 0.11801853 & 0.19991934 \\
  57.839 &        0.12 $\pm$ 0.23 &       1.79 $\pm$ 0.25 &        -1.34 $\pm$ 0.04 &      -31.93 $\pm$ 0.08 & 342.412 $\pm$ 0.2 &  -3.4 $\pm$ 0.007751 & 0.16773936 & 0.11801885 & 0.19991972 \\
  61.689 &        1.13 $\pm$ 0.23 &       1.36 $\pm$ 0.25 &        -0.15 $\pm$ 0.04 &      -30.07 $\pm$ 0.08 & 342.455 $\pm$ 0.2 &  -2.9 $\pm$ 0.007743 & 0.17031014 & 0.11823253 & 0.20058017 \\
  61.694 &        1.12 $\pm$ 0.23 &       1.78 $\pm$ 0.25 &         1.71 $\pm$ 0.04 &      -29.42 $\pm$ 0.08 & 342.444 $\pm$ 0.2 &  -2.9 $\pm$ 0.007743 & 0.17031263 & 0.11823256 & 0.20058108 \\
  64.828 &        0.55 $\pm$ 0.23 &       1.31 $\pm$ 0.25 &        -2.88 $\pm$ 0.04 &      -30.52 $\pm$ 0.08 & 342.511 $\pm$ 0.2 &  -3.8 $\pm$ 0.007775 & 0.17173754 &  0.1186557 & 0.20059256 \\
   64.83 &         0.0 $\pm$ 0.23 &       0.62 $\pm$ 0.25 &         -1.1 $\pm$ 0.04 &       -31.2 $\pm$ 0.08 & 342.566 $\pm$ 0.2 &  -3.8 $\pm$ 0.007775 & 0.17173871 & 0.11865608 & 0.20059269 \\
  68.649 &        2.39 $\pm$ 0.22 &       1.98 $\pm$ 0.24 &        -0.59 $\pm$ 0.04 &      -29.52 $\pm$ 0.08 & 342.324 $\pm$ 0.2 & -2.37 $\pm$ 0.007565 & 0.17253717 & 0.11883309 & 0.20170115 \\
  68.651 &        1.92 $\pm$ 0.22 &        1.5 $\pm$ 0.24 &          2.0 $\pm$ 0.04 &      -28.54 $\pm$ 0.08 & 342.306 $\pm$ 0.2 & -2.36 $\pm$ 0.007565 & 0.17253756 & 0.11883319 & 0.20170174 \\
  83.789 &       -2.79 $\pm$ 0.22 &      -0.95 $\pm$ 0.26 &        -4.49 $\pm$ 0.04 &      -30.45 $\pm$ 0.08 & 343.011 $\pm$ 0.2 & -6.64 $\pm$ 0.007611 & 0.16181003 & 0.11692285 & 0.19933288 \\
  83.791 &       -2.77 $\pm$ 0.22 &      -1.68 $\pm$ 0.26 &        -3.58 $\pm$ 0.04 &      -30.34 $\pm$ 0.08 & 342.992 $\pm$ 0.2 & -6.64 $\pm$ 0.007611 & 0.16180857 & 0.11692261 & 0.19933255 \\
  87.757 &       -2.13 $\pm$ 0.23 &      -0.49 $\pm$ 0.26 &         1.12 $\pm$ 0.04 &      -31.72 $\pm$ 0.08 & 342.741 $\pm$ 0.2 & -5.56 $\pm$ 0.007656 &    0.16288 & 0.11703363 & 0.19997583 \\
   87.76 &       -1.98 $\pm$ 0.23 &      -0.57 $\pm$ 0.26 &         -0.6 $\pm$ 0.04 &      -32.46 $\pm$ 0.08 & 342.711 $\pm$ 0.2 & -5.55 $\pm$ 0.007656 & 0.16288094 & 0.11703379 & 0.19997624 \\
  89.758 &       -1.86 $\pm$ 0.23 &       0.08 $\pm$ 0.25 &         1.09 $\pm$ 0.04 &      -32.45 $\pm$ 0.08 & 342.599 $\pm$ 0.2 & -5.29 $\pm$ 0.007789 & 0.16464062 & 0.11732348 & 0.20017845 \\
  89.765 &        -1.4 $\pm$ 0.23 &      -0.43 $\pm$ 0.25 &         0.07 $\pm$ 0.04 &       -31.8 $\pm$ 0.08 & 342.599 $\pm$ 0.2 & -5.29 $\pm$ 0.007789 & 0.16464785 &  0.1173249 & 0.20017932 \\
  93.734 &        2.22 $\pm$ 0.23 &        1.9 $\pm$ 0.25 &         3.09 $\pm$ 0.04 &      -28.12 $\pm$ 0.08 & 342.239 $\pm$ 0.2 &  -2.4 $\pm$ 0.007821 & 0.16667385 & 0.11748946 & 0.20113443 \\
  93.737 &        2.32 $\pm$ 0.22 &       2.36 $\pm$ 0.25 &         3.42 $\pm$ 0.04 &      -29.61 $\pm$ 0.08 & 342.239 $\pm$ 0.2 & -2.39 $\pm$ 0.007821 &  0.1666749 &  0.1174897 & 0.20113499 \\
  96.701 &         0.6 $\pm$ 0.23 &       1.24 $\pm$ 0.24 &         2.46 $\pm$ 0.04 &      -28.81 $\pm$ 0.08 & 342.286 $\pm$ 0.2 & -3.49 $\pm$ 0.007804 & 0.16715045 & 0.11760885 & 0.20136897 \\
  96.703 &        1.29 $\pm$ 0.23 &       0.81 $\pm$ 0.25 &         3.15 $\pm$ 0.04 &      -27.72 $\pm$ 0.08 & 342.361 $\pm$ 0.2 & -3.49 $\pm$ 0.007804 & 0.16715089 & 0.11760895 & 0.20136918 \\
 102.754 &        2.26 $\pm$ 0.25 &       4.06 $\pm$ 0.26 &          4.5 $\pm$ 0.04 &      -30.67 $\pm$ 0.08 & 342.174 $\pm$ 0.2 & -3.14 $\pm$ 0.008778 & 0.16976136 &  0.1186475 & 0.20205192 \\
 110.683 &        3.69 $\pm$ 0.23 &       2.88 $\pm$ 0.23 &         9.94 $\pm$ 0.04 &      -28.11 $\pm$ 0.08 & 341.805 $\pm$ 0.2 & -0.58 $\pm$ 0.007984 & 0.17549463 &  0.1203263 & 0.20394538 \\
 110.686 &        3.73 $\pm$ 0.23 &       2.76 $\pm$ 0.23 &        10.16 $\pm$ 0.04 &      -26.96 $\pm$ 0.08 & 341.718 $\pm$ 0.2 & -0.58 $\pm$ 0.007983 & 0.17549681 & 0.12032694 &  0.2039461 \\
 120.631 &       -0.48 $\pm$ 0.23 &      -2.07 $\pm$ 0.24 &          7.4 $\pm$ 0.04 &      -27.72 $\pm$ 0.08 & 342.092 $\pm$ 0.2 &  -2.6 $\pm$ 0.007746 &  0.1730541 &  0.1193655 & 0.20294897 \\
 123.656 &       -0.41 $\pm$ 0.23 &      -1.49 $\pm$ 0.24 &         8.13 $\pm$ 0.04 &      -28.93 $\pm$ 0.08 & 342.117 $\pm$ 0.2 & -2.77 $\pm$ 0.007689 & 0.17193214 &   0.118814 & 0.20271946 \\
 125.635 &       -0.77 $\pm$ 0.23 &      -2.23 $\pm$ 0.24 &         3.71 $\pm$ 0.04 &      -26.53 $\pm$ 0.08 & 342.211 $\pm$ 0.2 & -3.51 $\pm$ 0.007703 & 0.17092289 & 0.11899337 & 0.20230043 \\
 128.678 &       -2.53 $\pm$ 0.25 &      -3.91 $\pm$ 0.26 &         1.76 $\pm$ 0.04 &      -28.74 $\pm$ 0.08 & 342.368 $\pm$ 0.2 &  -4.3 $\pm$ 0.008513 & 0.17089052 & 0.11899774 &  0.2023141 \\
  131.67 &       -1.08 $\pm$ 0.23 &      -2.52 $\pm$ 0.25 &        -2.25 $\pm$ 0.04 &      -28.69 $\pm$ 0.08 & 342.465 $\pm$ 0.2 & -3.81 $\pm$ 0.007745 & 0.16833018 & 0.11833214 & 0.20184515 \\
 137.594 &        1.77 $\pm$ 0.23 &      -0.31 $\pm$ 0.25 &        -5.83 $\pm$ 0.04 &      -27.91 $\pm$ 0.08 & 342.496 $\pm$ 0.2 & -2.74 $\pm$ 0.007738 & 0.16802667 & 0.11829976 & 0.20146651 \\
 137.599 &        1.66 $\pm$ 0.23 &      -0.53 $\pm$ 0.25 &         -5.2 $\pm$ 0.04 &       -27.5 $\pm$ 0.08 & 342.527 $\pm$ 0.2 &  -3.4 $\pm$ 0.007773 & 0.16761446 & 0.11833745 & 0.20142893 \\
 140.621 &        0.82 $\pm$ 0.23 &      -0.78 $\pm$ 0.26 &        -9.17 $\pm$ 0.04 &      -28.78 $\pm$ 0.08 & 342.529 $\pm$ 0.2 &  -3.9 $\pm$ 0.007811 & 0.16689975 & 0.11811073 &  0.2015046 \\
 140.624 &         1.0 $\pm$ 0.23 &      -0.71 $\pm$ 0.26 &        -8.29 $\pm$ 0.04 &      -28.75 $\pm$ 0.08 & 342.487 $\pm$ 0.2 & -3.91 $\pm$ 0.007772 & 0.16699448 & 0.11812297 & 0.20143423 \\
 151.601 &        3.81 $\pm$ 0.23 &       1.25 $\pm$ 0.23 &         -2.8 $\pm$ 0.04 &      -28.08 $\pm$ 0.08 & 341.776 $\pm$ 0.2 & -0.12 $\pm$ 0.007762 & 0.17500772 &  0.1197036 & 0.20387995 \\

\enddata
\end{deluxetable}

\end{document}